\documentclass[12pt]{iopart}

\expandafter\let\csname equation*\endcsname\relax
\expandafter\let\csname endequation*\endcsname\relax
\usepackage{amsmath}
\usepackage{iopams}
\usepackage{bbm}
\usepackage{amsfonts}
\usepackage{amssymb}
\usepackage{mathtools}
\usepackage{thmtools}
\usepackage{hyperref}
\usepackage[nameinlink,noabbrev]{cleveref}
\usepackage{xcolor}
\usepackage{nicefrac}
\usepackage[page,toc]{appendix}
\usepackage{tikz}
\usepackage[normalem]{ulem}
\usepackage{enumitem}
\usepackage[font=footnotesize,labelfont=bf]{caption}
\usepackage{subcaption}
\usepackage{tabularx}

\usetikzlibrary{shapes,patterns,patterns.meta}
\tikzstyle{block} = [rectangle, draw, fill=gray!20, text width=2em, text centered, rounded corners, minimum height=2em]

\newtheorem{theorem}{Theorem}
\newtheorem{statement}{Statement}

\crefname{subsection}{subsection}{subsections}
\crefname{statement}{statement}{statements}

\DeclarePairedDelimiter{\bra}{\langle}{\rvert}%
\DeclarePairedDelimiter{\ket}{\lvert}{\rangle}%
\DeclarePairedDelimiterX\innerp[2]{\langle}{\rangle}{#1\delimsize\vert\mathopen{}#2}%
\DeclarePairedDelimiterX\braket[2]{\langle}{\rangle}{#1\delimsize\vert\mathopen{}#2}%
\DeclarePairedDelimiterX\braketOP[3]{\langle}{\rangle}{#1\,\delimsize\vert\,\mathopen{}#2\,\delimsize\vert\,\mathopen{}#3}%
\DeclarePairedDelimiterX\ketbra[2]{\lvert}{\rvert}{#1\delimsize\rangle\!\delimsize\langle#2}%
\DeclarePairedDelimiterX\outerp[2]{\lvert}{\rvert}{#1\delimsize\rangle\!\delimsize\langle#2}%
\DeclarePairedDelimiterX\projector[1]{\lvert}{\rvert}{#1\delimsize\rangle\!\delimsize\langle#1}%
\DeclarePairedDelimiter\abs{\lvert}{\rvert}%

\newcommand{\bbPi}{\hspace{0.6em}\raisebox{0.17em}{\scalebox{1}[0.65]{\(\vert\)}}\hspace{-0.65em}\raisebox{0.17em}{\scalebox{1}[0.65]{\(\vert\)}}\hspace{-0.43em}\Pi}
\newcommand\mydots{\hbox to 1em{.\hss.\hss.}}

\makeatletter
\newrobustcmd{\fixappendix}{%
	\patchcmd{\l@section}{1.5em}{7em}{}{}%
	\patchcmd{\l@subsection}{2.3em}{7em}{}{}%
}
\makeatother

\begin{document}
	\title[Correlation as a Resource in Unitary Quantum Measurements]{Correlation as a Resource in Unitary Quantum Measurements}
	\author[]{\parbox{\linewidth}{Vishal Johnson\(^{1,2}\), Ashmeet Singh\(^{3}\), Reimar Leike, Philipp Frank\(^{1}\), Torsten En{\ss}lin\(^{1,4,2,5}\)}}
	
	\address{\(^1\) Max-Planck-Institut f{\"u}r Astrophysik, Karl-Schwarzschild-Stra{\ss}e 1, 85748 Garching, Germany.}
	\address{\(^2\) Ludwig-Maximilians-Universit{\"a}t,  Geschwister-Scholl-Platz 1, 80539 M{\"u}nchen, Germany.}
	\address{\(^3\) Indian Institute of Technology Delhi, Hauz Khas, New Delhi, 110016, India.}
	\address{\(^4\) Deutsches Zentrum f{\"u}r Astrophysik, Postplatz 1, 02826 G{\"o}rlitz, Germany.}
	\address{\(^5\) Excellence Cluster ORIGINS, Boltzmannstr. 2, 85748 Garching, Germany.}
	\ead{vishal@mpa-garching.mpg.de, ashmeet@iitd.ac.in, reimar@leike.name, phfrank@stanford.edu, ensslin@mpa-garching.mpg.de}
	
	\begin{abstract}
		Quantum measurement is a physical process. What physical resources and constraints does quantum mechanics require for measurement to produce the classical world we observe? Treating measurement as a fully unitary quantum process, our goal is to show that objective, redundant, and correctly aligned outcomes are possible iff the environment begins in a specially structured, correlated subspace. We start with a minimal set of assumptions: unitarity, orthogonality of conditional environment branches, and finite-dimensional Hilbert spaces. Using these, we demonstrate that generic environmental states cannot support redundant and mutually consistent records of the signal, the measured quantum system. The admissible initial states form a subspace on which the measurement maps obey the Knill-Laflamme error-correction conditions, revealing that the emergence of classical objectivity relies on the environment behaving like a quantum error-correcting code. The post-measurement subspace naturally factorizes into a ``pointer'' to hold measurement outcomes and ``memory'' to retain pre-measurement quantum information about the environment's state, thereby respecting the no-deletion theorem. This further allows the identification of correlation as a finite resource consumed during measurement. Through an explicit qudit model with local interactions, we demonstrate how correlated environments yield redundant observer networks. Simulations show that record fidelity and redundancy depend on the initial correlations in the environment. This perspective links quantum Darwinism to error correction and raises the possibility that natural processes may prepare and evolutionarily favour environments capable of supporting reliable measurement.
	\end{abstract}
	\textbf{keywords:} unitary quantum mechanics, objective classical reality, quantum Darwinism, 
	observer network state, correlation, quantum error correction
	
	\section{Introduction: Measurement in a Unitary World} \label{sec:intro}
	In classical physics, measurement involves probing a system to determine its pre-existing state. We represent its state by a point in phase space which, for instance, needs specification of \(6N\) real numbers describing the position and momenta of \(N\) particles in three spatial dimensions. Crucially, the state of the system is assumed to exist independently of the observer and the measurement procedure, and the act of observation merely serves to extract objective information about the system without disturbing it. On the other hand, the notion of measurement in quantum mechanics is more subtle. Unlike in classical physics, we cannot treat the observer as a passive participant; it plays an essential dynamical role in the process. For example, in the standard decoherence-based description \cite{PhysRevD.26.1862, PhysRevD.24.1516, 1985ZPhyB..59..223J, Zurek_2000} (\cite{SCHLOSSHAUER20191, RevModPhys.76.1267, RevModPhys.75.715} for more recent reviews) of quantum measurement \cite{Busch1991, tomaz2025quantummeasurementproblemreview}, the measured system (henceforth called \emph{signal}) and corresponding environment are treated as parts of a single closed quantum system evolving unitarily. Through their mutual interaction, the signal becomes entangled \cite{zurek-classical} with distinct states of the environment that correlate with particular pointer states of the signal. As these environmental states evolve, they rapidly become orthogonal to one another, effectively suppressing interference between different outcomes and leading to the emergence of classicality.
	
	Quantum Darwinism \cite{Zurek_2000, Zurek_2009, zurek-2022} complements this approach, emphasizing that classical reality arises not merely from decoherence, but from the formation of redundant and robust records of the signal state across many independent fragments of the environment \cite{PhysRevLett.93.220401}. These records allow multiple observers to infer the same outcome without disturbing the signal, thereby providing an operational basis for objective classical reality. However, this standard picture leaves an open question: are there any restrictions on the initial state of the environment for such redundancy and dynamic decoherence to occur in the first place? While decoherence and quantum Darwinism describe how classical behavior emerges given an appropriate environment, they typically assume that the environment a priori starts in a suitable ``ready'' state, unentangled with the signal \cite{Blume_Kohout_2005, PhysRevA.81.062110}. In this work, we investigate the structure required in the environment for the creation of distinct, redundant, and decohered records.
	
	To illustrate this point, we consider a toy example of a gedankenexperiment where a Stern-Gerlach apparatus measures the spin of an electron \cite{feynman-3}. As it is a spin \(\nicefrac{1}{2}\) particle, there would be two spots on a screen following the apparatus: one corresponding to the label spin up (0) and another to spin down (1) (see \cref{fig:stern-gerlach}). If we were to reverse the magnet systems comprising the Stern-Gerlach apparatus --- the previous North pole being the current South pole, and vice versa --- the spots on the screen would need to be labeled in the opposite order to maintain consistency with Stern-Gerlach apparatuses elsewhere. Thus, there is an environmental influence on the measurement procedure, and information about the environment must be present for quantum measurement to take place successfully. To consistently obtain correlated outcomes between the signal and the observer (a subset of the environment), information about the environment must be known or extracted.
	
	\begin{figure} [h!]
		\centering
		\begin{tikzpicture}
			\draw (-2,0) circle (0.05);
			\node at (-2,-0.5)  {electron};
			
			\draw (0,-1) rectangle ++(3,2);
			\node at (1.5,-1) {\phantom{Stern-Gerlach apparatus}};
			
			\draw plot [smooth] coordinates {(-2,0) (-1,0.1) (1.5,0.45) (4,0.5)};
			\draw plot [smooth] coordinates {(-2,0) (-1,-0.1) (1.5,-0.45) (4,-0.5)};
			
			\node at (1.5, 0) {\scalebox{1.5}{\(\mathbf{\uparrow}\)}};
			
			\draw (4,1) -- (4,-1);
			\node[anchor=west] at (4.2,0.5) {spin up (0)};
			\node[anchor=west] at (4.2,-0.5) {spin down (1)};
		\end{tikzpicture}
		\begin{tikzpicture}
			\draw (-2,0) circle (0.05);
			\node at (-2,-0.5)  {electron};
			
			\draw (0,-1) rectangle ++(3,2);
			\node at (1.5,-1.5) {Stern-Gerlach apparatus};
			
			\draw plot [smooth] coordinates {(-2,0) (-1,0.1) (1.5,0.45) (4,0.5)};
			\draw plot [smooth] coordinates {(-2,0) (-1,-0.1) (1.5,-0.45) (4,-0.5)};
			
			\node at (1.5, 0) {\scalebox{1.5}{\(\mathbf{\downarrow}\)}};
			
			\draw (4,1) -- (4,-1);
			\node[anchor=west] at (4.2,0.5) {spin down (1)};
			\node[anchor=west] at (4.2,-0.5) {spin up (0)};
		\end{tikzpicture}
		\caption{\label{fig:stern-gerlach} An electron is a spin \nicefrac{1}{2} particle. The passage of several electrons through an appropriately prepared Stern-Gerlach apparatus (adjusted for the electron charge) results in two spots on a fluorescent screen. One spot would correspond to spin up (0) and another to spin down (1) states of the electrons. The large arrow (\(\mathbf{\uparrow}/\mathbf{\downarrow}\)) indicates the magnet configuration of the Stern-Gerlach apparatus that causes state separation. If the configuration of the Stern-Gerlach magnets is reversed, so have to be the labels associated with the spots on the screen corresponding to spin up and spin down. The measurement apparatus (environment) affects the measurement outcomes, and we have to correct for this.}
	\end{figure}
	
	We elaborate this example in the language of quantum mechanics. First, we split the global Hilbert space as follows: the electron spin is taken as the ``signal \(S\)'' spanned by a 2-dimensional Hilbert space, \(\mathcal{H}_{S} \simeq \text{span}\{\ket{0}_{S},\ket{1}_{S}\}\). We take the Stern-Gerlach magnet as ``apparatus A'' which for simplicity we model as a coarse-grained magnet with two possible orientations: \(\mathcal{H}_{A} \simeq \text{span}\{\ket{\uparrow}_{A},\ket{\downarrow}_{A}\}\), with \(\ket{\uparrow}_{A}\) representing a particular North-South orientation of the magnetic poles, and \(\ket{\downarrow}_{A}\) representing the reversed orientation. The ``observer O'' is taken as the part of the screen which records and registers the measurement outcome and is described by \(\mathcal{H}_{O} \simeq \text{span}\{\ket{0}_{O},\ket{1}_{O}\}\): conditional observer states corresponding to the electron spin measurement outcome. The apparatus and observer are considered to be effectively classicalised systems with many degrees of freedom; the above is supposed only to be a coarse-grained description. For example, the magnets that comprise the Stern-Gerlach apparatus could correspond to very many magnetic domains pointing \(\uparrow\) or \(\downarrow\). Towards the end of this section, we fine-grain our description slightly to demonstrate a consistent unitary measurement procedure.
	
	As in thermodynamics and in agreement with standard decoherence literature, we consider the environment to encompass everything else relevant to the measurement procedure, other than the signal itself, such as observers, apparatuses, photon baths, etc., that interact with the signal, among others. Thus, in our example, we take the ``environment'' to be the combination of observer and apparatus, \(\mathcal{H}_{E} \simeq \mathcal{H}_{O} \otimes \mathcal{H}_{A}\). The global Hilbert space is \(\mathcal{H} \simeq \mathcal{H}_{S} \otimes \mathcal{H}_{O} \otimes \mathcal{H}_{A}\). We model both, the observer and apparatus, to begin in an initially unentangled state with each other and with the signal, making the initial state a product state:
	\begin{equation}
		\ket*{\Psi^\text{init}} = \ket{\psi}_{S} \otimes \ket{\chi}_{O} \otimes \ket{\phi}_{A}.
	\end{equation}
	
	We take arbitrary (normalised) initial states of the signal and observer: \(\ket{\psi}_{S} = \psi_{0}\ket{0}_{S} + \psi_{1}\ket{1}_{S}\) and \(\ket{\chi}_{O} = \chi_{0}\ket{0}_{O} + \chi_{1}\ket{1}_{O}\). For the Stern-Gerlach apparatus, let us consider the initial state \(\ket{\phi}_{A} = \ket{\uparrow}_{A}\). We propose that there exists a specific unitary transformation that leads to the final post-measurement state, referred to as \(\ket{\Psi^\text{final}}\) (we show such an explicit unitary transformation in \cref{sec:meas_proc} for a generalised version of the discussion here):
	\begin{equation} \label{eq:stern_gerlach_correct}
		\ket*{\Psi^\text{final}_{_{{A=\uparrow}}}} = \Big(\psi_{0}\ket{0}_{S}\ket{0}_{O} + \psi_{1}\ket{1}_{S}\ket{1}_{O} \Big) \otimes\Big(\chi_0\ket{\uparrow}_{A}  + \chi_1\ket{\downarrow}_{A} \Big).
	\end{equation}
	The above state represents the observer entangled and correctly correlated with distinct pointer states of the signal, with coefficients matching what one would expect from the Born rule. Further, as a consequence of unitarity, information must be preserved \cite{no-cloning,no-deletion}, thereby leading to the initial state of the observer being absorbed by the rest of the environment as seen in \cref{eq:stern_gerlach_correct}. It, thus, seems like the state of the magnet is being flipped by the observer. However, as would be clarified later in this section in \cref{eq:stern_gerlach_make_work}, it is only one qubit of information about the magnet’s state that is flipped.
	
	Another important consequence of unitarity is evident if we start the Stern-Gerlach apparatus in a state with the magnet poles reversed, that is, \(\ket{\phi}_{A} = \ket{\downarrow}_{A}\) state. In this case, unitarity demands that the post-measurement outcomes map to an orthogonal subspace:
	\begin{equation}
		\ket*{\Psi^\text{final}_{_{{A=\downarrow}}}} = \Big(\psi_{0}\ket{0}_{S}\ket{1}_{O} + \psi_{1}\ket{1}_{S}\ket{0}_{O} \Big) \otimes \Big(\chi_0\ket{\uparrow}_{A}  + \chi_1\ket{\downarrow}_{A} \Big).
	\end{equation}
	The above state corresponds to the observer being anti-aligned with the signal pointer states. In either case, the reduced density matrix of the \emph{observer} post-measurement is diagonal, representing effective decoherence,
	\begin{equation} \label{eq:stern_gerlach_correct_density}
		\rho^\text{final}_{{O}_{{A=\uparrow}}} = \tr_{S,A}\bigg(\ketbra*{\Psi^\text{final}_{_{{A=\uparrow}}}}{\Psi^\text{final}_{_{{A=\uparrow}}}}\bigg) = \abs{\psi_0}^{2} \ketbra{0}{0}_{O} + \abs{\psi_1}^{2} \ketbra{1}{1}_{O} ,
	\end{equation}
	and
	\begin{equation} 
		\rho^\text{final}_{{O}_{{A=\downarrow}}} = \tr_{S,A}\bigg(\ketbra*{\Psi^\text{final}_{_{{A=\downarrow}}}}{\Psi^\text{final}_{_{{A=\downarrow}}}}\bigg) = \abs{\psi_1}^{2} \ketbra{0}{0}_{O} + \abs{\psi_0}^{2} \ketbra{1}{1}_{O}.
	\end{equation} 
	However, in \(\rho^\text{final}_{{O}_{{A=\downarrow}}}\), as a result of the initial state of the environment being different, there is misalignment in terms of probabilities assigned to various outcomes using the Born rule. Physically, the spots corresponding to labels 0 and 1 would have opposite intensities. In terms of labels, this is akin to exchanging the labels associated with the outcomes of the observer, as shown in \cref{fig:stern-gerlach}.
	
	This environmental influence is further pronounced if we look at an arbitrary superposition of the Stern-Gerlach apparatus, \(\ket{\phi}_{A} = \phi_{\uparrow}\ket{\uparrow}_{A} + \phi_{\downarrow}\ket{\downarrow}_{A}\), which will result in
	\begin{multline} \label{eq:stern_gerlach_dont_work}
		\ket*{\Psi^\text{final}} = \bigg[\phi_{\uparrow}\Big(\psi_{0}\ket{0}_{S}\ket{0}_{O} + \psi_{1}\ket{1}_{S}\ket{1}_{O} \Big)  + \phi_{\downarrow}\Big(\psi_{0}\ket{0}_{S}\ket{1}_{O} + \psi_{1}\ket{1}_{S}\ket{0}_{O} \Big) \bigg] \\
		\otimes \Big(\chi_0\ket{\uparrow}_{A}  + \chi_1\ket{\downarrow}_{A} \Big),
	\end{multline}
	for which the reduced density matrix of the observer, written in the \(\{\ket{0}_{O},\ket{1}_{O}\}\) basis, takes the form
	\begin{equation} \label{eq:stern_gerlach_dont_work-density}
		\rho_{O}^\text{final} = \begin{alignedat}{4}
			\Big(\abs{\phi_{\uparrow}}^2\abs{\psi_0}^2  &+ \abs{\phi_{\downarrow}}^2\abs{\psi_1}^2 \Big) & &\ketbra{0}{0}_{O}
			& +\Big(\phi_\uparrow\phi^*_\downarrow \abs{\psi_0}^2 &+ \phi^*_\uparrow\phi_\downarrow \abs{\psi_1}^2\Big) & &\ketbra{0}{1}_{O}
			\\
			+\Big(\phi^*_\uparrow\phi_\downarrow \abs{\psi_0}^2 &+ \phi_\uparrow\phi^*_\downarrow \abs{\psi_1}^2\Big) & &\ketbra{1}{0}_{O}
			& +\Big(\abs{\phi_{\downarrow}}^2\abs{\psi_0}^2 &+ \abs{\phi_{\uparrow}}^2\abs{\psi_1}^2\Big) & &\ketbra{1}{1}_{O}
		\end{alignedat}.
	\end{equation}
	This is generally \emph{not} diagonal. We can make a similar argument at the level of the signal density matrix, illustrating that the initial environmental state plays a crucial role in determining whether the observer decoheres and correctly correlates with the measured system. This simple model illustrates that unitarity alone dictates a non-trivial dependence of the measurement outcome on the initial configuration of the environment. We elaborate on the structure of this special environmental subspace in this work, contingent on some broad requirements for what qualifies as objective classical reality.
	
	To formalize and guide our notion of objective classical reality and connect it more closely to decoherence and quantum Darwinism, we utilize \emph{spectrum broadcast structures} \cite{spectrum-broadcast-0,spectrum-broadcast-1}(SBSs). In the spirit of the works on SBSs, we assume that the state of a system \(S\) exists objectively if multiple observers can probe the signal independently and without perturbing it. As before, we split the Hilbert space into signal \(\mathcal{H}_{S}\) and environment \(\mathcal{H}_{E}\), with the environment further consisting of many smaller fragments (which could include the multiple observers probing the signal),
	\begin{equation}
		\mathcal{H} \simeq \mathcal{H}_{S} \otimes \mathcal{H}_{E} \simeq \mathcal{H}_{S} \otimes \Big(\mathcal{H}_{E_1} \otimes \mathcal{H}_{E_2}  \cdots \otimes \mathcal{H}_{{E}_N}\Big).
	\end{equation}
	Horodecki et al. \cite{spectrum-broadcast-0} posit that at late time, we can write the global density matrix of the combined signal-environment (for a fraction \(fE\) of the environment) in the special form corresponding to measurement resulting in objective classical outcomes:
	\begin{equation} \label{eq:sbs}
		\rho_{{S}:f{E}}^\text{final} =  \sum_{i}\bbPi_i\ketbra{\psi}{\psi}_{S}\bbPi^\dagger_i \otimes \bigotimes_{\alpha\in fE}\rho^i_\alpha.
	\end{equation}
	Here \(\{\bbPi_i\}\) are projectors on the Hilbert space of the signal, one for each measurement outcome \(i\), and \(\{\bbPi_i\ket{\psi}_{S}\}\) are the relevant states of the signal corresponding to pointer \(i\) which the environment records. Each pointer has a Born probability of \(p_i=\braketOP{\psi}{\bbPi_i}{\psi}_{S}\) (with \(\sum_i p_{i} = 1\)), and \(\rho^i_\alpha\) is the conditional reduced density matrix of the \(\alpha\) fragment of the environment \(\mathcal{H}_{E_{\alpha}}\) corresponding to the \(i\)-th signal outcome. We use the Born rule (see \cite{PhysRevLett.90.120404} for the context of the Born rule within the decoherence paradigm) without justification in this paper; however, its interpretation as a probability is of no relevance to the results discussed here.
	
	In addition to the signal density matrix \(\rho_{S}^\text{final} = \tr_{E}\big(\rho^\text{final}\big)\) to be diagonal in the pointer basis, the structure of an SBS state brings forth some important properties of objective classical reality:
	\begin{itemize}
		\item \emph{Distinguishable records:} The environmental records corresponding to distinct system outcomes must be \emph{perfectly distinguishable}, ensuring that each fragment \(E_\alpha\) carries unambiguous classical information,
		\begin{equation}
			\rho^i_\alpha \rho^j_\alpha = 0 \quad \forall i \neq j.
		\end{equation}	
		\item \emph{Redundancy and strong independence:} Objectivity requires not only distinguishability but redundant accessibility. Each environmental fragment \(E_\alpha\) must store an independent imprint of the measurement process. We formalise this notion through \emph{strong independence:} conditioned on the signal’s outcome, distinct fragments are uncorrelated. Mathematically, this says that the mutual information\footnote{For quantum systems, the von-Neumann entropy for system \(A\) is defined as \(S(\rho_{A}) = \tr(\rho_{A}\log(\rho_{A}))\). For a combined system \(A \otimes B\), the mutual information is defined as \(I({A:B}) = S(\rho_{A}) + S(\rho_{B}) - S(\rho_{AB})\). Further details can be found in \cite{nielsen-chuang}.} \(I\) between fragments \({E_\alpha}\) and \({E_\beta}\) vanishes for \(\alpha \neq \beta\) if conditioned on the signal \(S\),
		\begin{equation}
			I({E_\alpha}:{E_\beta}\vert{S}) = 0 \quad \forall \alpha\neq\beta.
		\end{equation}
		The above equation further implies that each fragment contains redundantly encoded records in the spirit of quantum Darwinism.
		\item \emph{Correct alignment:} Finally, each fragment of the environment must be correctly aligned with the pointer state of the signal. If the correlations are misaligned (as in the reversed Stern-Gerlach example discussed earlier), the recorded outcomes become inconsistent, and the shared classical reality among observers breaks down. Correct alignment guarantees that every observer reading from different parts of the environment agrees on the same outcome, faithfully reflecting the signal’s state in the pointer basis.
	\end{itemize}
	
	If an environment is redundantly encoded as suggested by objective reality, it contains information about itself. Thus, its influence on the measurement procedure can be corrected for, enabling successful quantum measurement even for more general initial apparatus states. We illustrate this by continuing the Stern-Gerlach example introduced before. As discussed before, for the measurement unitary being considered, the arbitrary state \(\ket{\phi}_{A} = \phi_{\uparrow}\ket{\uparrow}_{A} + \phi_{\downarrow}\ket{\downarrow}_{A}\) does not lead to successful measurement. Let us consider the same apparatus but this time encoded redundantly; \(\ket{\phi}_{A} = \phi_\uparrow\ket{\uparrow}_{A_1}\ket{\uparrow\uparrow\dots}_{A_2} + \phi_\downarrow\ket{\downarrow}_{A_1}\ket{\downarrow\downarrow\dots}_{A_2}\) with \(A = {A_1}\otimes{A_2}\) being a relevant subdivision. As the notation indicates, \(A_2\) corresponds to the physical state of the magnet consisting of many magnetic domains, and \(A_1\) contains a record of the state of the physical magnet \(A_2\). \(A_1 \otimes A_2\) are in an entangled state that share a common ``classical reality''. ``Knowing'' the state of the magnet allows one to correct for it and result in a consistent quantum measurement. In this case, a unitary operation
	\[\begin{aligned}
		\ket{\uparrow}_{A_1}\ket{\uparrow\uparrow\dots}_{A_2} &\to \ket{\uparrow}_{A_1}\ket{\uparrow\uparrow\dots}_{A_2}, & \quad \ket{\downarrow}_{A_1}\ket{\downarrow\downarrow\dots}_{A_2} &\to \ket{\uparrow}_{A_1}\ket{\downarrow\downarrow\dots}_{A_2}, \\
		\ket{\downarrow}_{A_1}\ket{\uparrow\uparrow\dots}_{A_2} &\to \ket{\downarrow}_{A_1}\ket{\uparrow\uparrow\dots}_{A_2}, & \ket{\uparrow}_{A_1}\ket{\downarrow\downarrow\dots}_{A_2} &\to \ket{\downarrow}_{A_1}\ket{\downarrow\downarrow\dots}_{A_2},
	\end{aligned}\]
	can be used to perform ``environmental correction'' to bring the apparatus to the state \(\ket{\phi}_{A} = \ket{\uparrow}_{A_1}\otimes\big(\phi_\uparrow\ket{\uparrow\uparrow\dots}_{A_2} + \phi_\downarrow\ket{\downarrow\downarrow\dots}_{A_2}\big)\) after which the first apparatus qubit \({A_1}\) can be used to obtain a consistent measurement result following \cref{eq:stern_gerlach_correct}:
	\begin{multline} \label{eq:stern_gerlach_make_work}
		\ket{\psi}_{S} \otimes \ket{\chi}_{O} \otimes \Big(\phi_\uparrow\ket{\uparrow}_{A_1}\ket{\uparrow\uparrow\dots}_{A_2}  + \phi_\downarrow\ket{\downarrow}_{A_1}\ket{\downarrow\downarrow\dots}_{A_2}\Big) \to \\
		\ket*{\Psi^\text{final}} = \Big(\psi_{0}\ket{0}_{S}\ket{0}_{O} + \psi_{1}\ket{1}_{S}\ket{1}_{O}\Big) \otimes \ket{\chi}_{A_1} \otimes \Big(\phi_\uparrow\ket{\uparrow\uparrow\dots}_{A_2}  + \phi_\downarrow\ket{\downarrow\downarrow\dots}_{A_2}\Big).
	\end{multline}
	The observer density matrix is \(\abs{\psi_0}^{2} \ketbra{0}{0}_{O} + \abs{\psi_1}^{2} \ketbra{1}{1}_{O}\) as in \cref{eq:stern_gerlach_correct_density}.
	
	The takeaway is that a redundantly encoded environment gives rise to consistent measurement. In this picture, quantum measurement is simply a transfer of redundancy from the measurement device (which is a part of the environment) to the observed signal measurement. The remainder of the paper elaborates on this idea. In \cref{sec:corr_res,sec:kl}, assuming a set of natural statements, we derive a no-go theorem that enables the sought-after interpretation of correlation as a resource. We demonstrate that realizing the requirements of objective classical reality under strictly unitary dynamics imposes stringent constraints on the initial environmental state: it must belong to a special correlated subspace that preserves inner products and ensures consistent correlations across observers. We provide a broad overview of the main theorem in \cref{sec:corr_res}, and some of the missing mathematical details are filled in in \cref{sec:kl}, where the proof of the main theorem concludes. On the way, in \cref{sec:kl}, we also make a formal connection to the Knill-Laflamme conditions in quantum error correction, with the environmental correlated subspace functioning as a code subspace that protects measurement consistency. Moreover, the demand for classicality induces a specific operator-algebraic structure on the environment Hilbert space --- one that factorizes into a pointer and a memory (of the prior environment). These constraints identify correlations as a finite resource enabling robust and redundant measurement outcomes. The results derived in these sections are generally applicable. In \cref{sec:meas_proc}, we develop a particular instance of a quantum measurement procedure to highlight these general ideas for a concrete instantiation. This also serves as a bridge to \cref{sec:num_res} where we corroborate our hypotheses through simulations. We conclude with a discussion in \cref{sec:discussion}.
	
	\section{Correlation as a Limited Resource: A No-Go Theorem} \label{sec:corr_res}
	The central result of this work is that correlation among environmental fragments constitutes a finite physical resource that enables quantum measurement. Here, correlation refers not merely to quantum entanglement, but to the consistent agreement between different parts of the environment about the outcome of a measurement. All fragments must encode the same classical information about the signal measurement. We demonstrate that objective classical reality, which requires distinguishable and redundant records, cannot emerge from arbitrary initial environmental states. A consistent measurement process is possible only when the environment begins within a special correlated subspace that supports the correct mapping between signal states and environmental records.
	
	To formalize this idea, we identify a minimal set of physically reasonable statements about the measurement process: \crefrange{stat:universality}{stat:arbitrary}. These statements are mutually incompatible; they cannot all be satisfied simultaneously. This incompatibility leads to a no-go theorem demonstrating that unitarity and objectivity jointly demand the environment to possess a special initial structure. In this section, we derive this no-go theorem, albeit with a few mathematical details which we fill in the next section. Furthermore, in the next section, we explore the mathematical structure of the special initial subspace for successful measurement, showing that it mirrors that of a quantum error-correcting code; it contains correlations that enable redundant measurement outcomes. We begin expressing the no-go theorem by listing the statements and commenting on them:
	\begin{statement}[Universality of Quantum Mechanics] \label{stat:universality}
		Unitary quantum mechanics holds universally.
	\end{statement}
	We assume that unitary quantum mechanics holds universally \cite{RevModPhys.29.454, Carroll2019}, including time evolution, which affects measurement-like processes by creating and decohering conditional signal-environment branches. We take our global Hilbert space to factorize between the system and environment, \(\mathcal{H} \simeq \mathcal{H}_{S} \otimes \mathcal{H}_{E}\). The global, pre-measurement signal-environment system begins in an \emph{unentangled} state,
	\begin{equation}
		\ket*{\Psi^\text{init}} = \ket{\psi}_{S}\otimes \ket{\phi}_{E},
	\end{equation}
	where \(\ket{\phi}_{E}\) is some initial state for the environment. The state evolves under a unitary map \(\mathcal{U}: \mathcal{H}_{S} \otimes \mathcal{H}_{E} \to \mathcal{H}_{S} \otimes \mathcal{H}_{E}\). A Hamiltonian description is often possible, \(\mathcal{U} = \exp(-iHt)\) with \(H \in \mathcal{L}(\mathcal{H})\) being the Hamiltonian and \(t\) the time coordinate in this specific reference frame. However, we stick to a description in terms of unitaries in this work:
	\begin{equation}
		\ket*{\Psi^\text{final}} = \mathcal{U}\big(\ket{\psi}_{S}\otimes \ket{\phi}_{E}\big).
	\end{equation}
	
	\begin{statement}[Quantum Measurement] \label{stat:measurement}
		We model quantum measurement through a unitary \(\mathcal{U}^\mathrm{meas}\) which maps the initial unentangled state of the signal-environment to a superposition of conditionally correlated and decohered branches,
		\begin{equation} \label{eq:quantum_ieasurement}
			\ket{\Psi^\mathrm{final}} = \mathcal{U}^\mathrm{meas}\ket{\psi}_{S}\ket{\phi}_{E} = \sum_i{ \Big(\bbPi_i\ket{\psi}_{S}\Big)  \otimes \ket{\mathcal{E}_{i\phi}}_{E}}.
		\end{equation}
	\end{statement}
	The state \(\ket{\mathcal{E}_{i\phi}}_{E}\) is the environmental state corresponding to the \(i\)-th pointer basis, where \(\{i\}\) label the different measurement outcomes. The state \(\ket{\mathcal{E}_{i\phi}}_{E}\) contains information about the initial state of the environment, \(\phi\), and a record of the measurement outcome, \(i\). \(\bbPi_i\) is the projector, on the signal subspace, for measurement outcome \(i\) with respect to the set of projective measurements, \(\{\bbPi_i\}\) on \(\mathcal{H}_{S}\). They are a set of orthonormal Hermitian projectors satisfying:
	\begin{subequations} \label{eq:orthonormal_projective_ieasurement}
		\begin{tabularx}{\textwidth}{XXX}
			\begin{equation} \label{eq:orthonormal_projective_ieasurement:hermitian}
				\bbPi_i^\dagger = \bbPi_i,
			\end{equation}
			&\begin{equation} \label{eq:orthonormal_projective_ieasurement:orthonormal}
				\bbPi_i\bbPi_j = \delta_{ij},
			\end{equation}
			&\begin{equation} \label{eq:orthonormal_projective_ieasurement:sum}
				\sum_i{\bbPi_i} = \mathbbm{1}_{S}.
			\end{equation}
		\end{tabularx}
	\end{subequations}

	We note that the measurement procedure described in \cref{eq:quantum_ieasurement} is a unitary action and no wavefunction collapse is involved. \Cref{eq:orthonormal_projective_ieasurement} above uses projective measurements \(\bbPi_i\) only as part of the definition of the unitary measurement procedure \cite{Baldijao2022quantumdarwinism, acevedo2024spectrumbroadcaststructuresvon}; we define it this way to be consistent with the empirical results of quantum measurements. Using projective valued measures, we assume sharp measurements. We do not consider a general result in terms of positive operator valued measures in this work, except for the comment that the Naimark dilation theorem \cite{nielsen-chuang} would allow our results to apply generally in a suitably chosen (larger than the signal) Hilbert space.
	
	The overall state of the signal and environment is therefore a superposition of projected states of the signal \(\bbPi_i\ket{\psi}_{S}\) entangled with environment states \(\ket{\mathcal{E}_{i\phi}}_{E}\) along the measurement degree of freedom \(i\). Such a measurement can be realized, for example, by an idealized von-Neumann scheme implemented through conditional unitaries \cite{ref1}
	\begin{equation}\label{eq:cond_unitary}
		\mathcal{U}^\text{meas} = \sum_{i} \bbPi_{i}\otimes \mathcal{U}_{i}.
	\end{equation}
	See \cref{sec:kl} for more details.
	
	\begin{statement}[Orthogonality of Conditional Environmental States] \label{stat:orthogonality}
		States of the environment corresponding to different measurement outcomes should be perfectly distinguishable, and therefore, orthogonal:
		\begin{equation}
			\braket{\mathcal{E}_{i\chi}}{\mathcal{E}_{j\phi}}_{E} = \delta_{ij} \braket{\mathcal{E}_{i\chi}}{\mathcal{E}_{i\phi}}_{E},
		\end{equation}
		for two different (not necessarily orthogonal) initial environmental states \(\ket{\phi}_{E}\) and \(\ket{\chi}_{E}\) and measurement outcomes \(i\) and \(j\).
	\end{statement}
	We demand this as a requirement for both effective decoherence of the system density matrix in the pointer basis, and to ensure that the environmental states provide reliable and distinguishable information about the measurement outcomes \cite{spectrum-broadcast-0,spectrum-broadcast-1}. We will see in \cref{sec:kl} that the consequences of demanding perfect orthogonality of conditional environmental states are two-fold:
	\begin{enumerate}[label=(\roman*)]
		\item inner products are preserved on the environmental states conditioned on the same outcome, \(\braket{\phi}{\chi}_{E} = \braket{\mathcal{E}_{i\chi}}{\mathcal{E}_{i\phi}}_{E}\), and,
		\item the initial environmental state cannot be arbitrary but has to belong to a proper subset \(\Phi\) of \(\mathcal{H}_{E}\), \(\ket{\phi}_{E} \in \Phi \subsetneq \mathcal{H}_{E}\).
	\end{enumerate}
	Measurement maps the initial environment state to distinct, orthogonal parts of the Hilbert space \(\Phi_i= \{\ket{\mathcal{E}_{i \phi}}_{E} \forall  \ket{\phi}_{E} \in \Phi\}\), \(\Phi_i\perp \Phi_j\) in case \(i \ne j\).
	
	\begin{statement}[Finite Dimensional Hilbert Space] \label{stat:finiteness}
		All Hilbert spaces involved are finite-dimensional.
	\end{statement}
	
	Most physical systems encountered in the laboratory possess a finite number of distinguishable outcomes and limited measurement resolution \cite{Busch1991}, leading naturally to an effectively finite-dimensional description. More fundamentally, insights from quantum gravity, such as black hole \cite{bekenstein1973} and holographic entropy bounds \cite{Bousso1999b, 'tHooft:1993gx,Susskind1995}, suggest that the Hilbert space of quantum gravity is locally finite-dimensional \cite{Bao_2017,banks2025finiteentropyimpliesfinite}. We therefore take finite dimensionality to be a physically well-motivated assumption and include it among the logical statements forming the basis of our no-go theorem. Finite-dimensionality limits the environment's ability to create multiple records of the measurement outcome. As discussed earlier, we take the environment \(\mathcal{H}_{E}\) to be further decomposed into various fragments, \(\mathcal{H} \simeq \bigotimes_{\alpha} \mathcal{H}_{E_\alpha}\). The measurement outcomes are labelled \(i\); let us call the number of such measurement outcomes as \(d\):
	\begin{equation} \label{eq:num_meas_out}
		d = \abs{\{i\}} = \text{number of measurement outcomes}.
	\end{equation}
	We label the number of redundant records of the signal measurement that are created in the environment as \(K: 0 \le K \leq N\). In \cref{sec:kl:multi_env}, we show that the dimension of the subspace of initial states, \(\Phi\), and the environmental Hilbert space, \(\mathcal{H}_{E}\), are constrained by the following inequality:
	\begin{equation} \label{eq:numerical_limits}
		K\ln(d) + \ln(\dim\Phi)\le \ln(\dim\mathcal{H}_{E}).
	\end{equation}
	A direct consequence of this is that for \(K\geq 1\) and \(\dim\mathcal{H}_{E} < \infty\), we find that \(\dim\Phi < \dim\mathcal{H}_{E}\).\footnote{An interesting feature is that the number of redundant classical records \(N\) increases as the logarithm of the dimension of the environment Hilbert space \(\dim\mathcal{H}_{E}\). This would balance out exactly the exponential growth of resources in quantum systems. It makes sense, as classical redundant records are the explicitly ``non-quantum'' aspect of quantum mechanics. There is no contradiction with quantum computation, where exponential (but fragile) quantum states are utilised for their exponential advantage, unlike in our case, where the linear (but resilient) ``classical'' records of quantum states are highlighted.} This means that the initial environment state cannot be an arbitrary one in the Hilbert space \(\mathcal{H}_{E}\) but must belong to a special subspace \(\Phi\) where \(\Phi\) is a strict subset of \(\mathcal{H}_{E}\). More redundant records imply a lower \(\dim\Phi\), which for multipartite environments implies that the initial state of the environment fragments has to be correlated. Therefore, assuming that \cref{eq:numerical_limits} holds, \crefrange{stat:universality}{stat:finiteness} imply that the following statement is false:
	\begin{statement}[Arbitrary Initial State] \label{stat:arbitrary}
		The initial state of the environment is arbitrary in the Hilbert space: \(\Phi = \mathcal{H}_{E}\).
	\end{statement}
	
	We, therefore, have \cref{th:no-go}:
	\begin{theorem}[Initial State No-Go Theorem] \label{th:no-go}
		\Crefrange{stat:universality}{stat:arbitrary} cannot be simultaneously true:
		\begin{enumerate}
			\item universality of unitary quantum mechanics,
			\item unitary quantum measurement procedure \cref{eq:quantum_ieasurement},
			\item orthogonality of environment states corresponding to different measurement outcomes,
			\item finite-dimensional Hilbert spaces, and,
			\item arbitrary initial state of the environment.
		\end{enumerate}
	\end{theorem}
	
	We have phrased \cref{th:no-go} as a no-go theorem so that the conditions going into its proof can be analysed in maximal generality. However, we take a specific stance on the statements: we give up \cref{stat:arbitrary}. Therefore, we take it that the initial state of the environment belongs to a strict subspace \(\Phi\) of the environment's physical Hilbert space \(\mathcal{H}_{E}\). The contradiction in \cref{th:no-go} arises because in a finite-dimensional environment, the available degrees of freedom in the environment constrain the number of orthogonal conditional subspaces required for redundant measurement. The environment is constrained to begin in a specially structured subspace. We progressively fill in the mathematical details of the proof of the theorem in the next section.
	
	\section{Constraining the Space of Initial Environment States} \label{sec:kl}
	We now examine the mathematical structure of both, the special subspace \(\Phi \subset \mathcal{H}_{E}\) of initial environment states and its map under \(\mathcal{U}_\text{meas}\). We show that the requirements of objective classical reality, specifically the orthogonality of conditional environmental states, impose constraints that are mathematically identical to the Knill-Laflamme (KL) conditions familiar from quantum error correction. Unitary evolution (for instance, the \(\mathcal{U}_{i}\) from \cref{eq:cond_unitary}) on this space of special initial states creates an isometric mapping to orthogonal subspaces corresponding to distinct outcomes. This further induces a natural decomposition of the post-measurement subspace of the environment space into ``pointer'' and ``memory'' (of the prior environment) components, capturing how system information is stored and protected, and ensuring consistency with the no-deletion theorem \cite{no-deletion} due to unitary evolution. In \cref{sec:kl:multi_env} we extend this discussion to multipartite environmental fragments, showing how the joint requirements of redundancy and KL-type structure give rise to the correlated pattern of records characteristic of objective classical reality.
	
	\subsection{Inner Product Preservation}\label{sec:kl:innerprod}
	We first demonstrate that the inner product of environmental states in \(\Phi\) is preserved under the unitary evolution modeling quantum measurement as mentioned in \cref{stat:measurement}. For two different initial states \(\ket{\phi}_{E}\) and \(\ket{\chi}_{E} \in \Phi\) for the environment we get
	\begin{equation} 
		\bra{\psi}_{S}\bra{\chi}_{E}\,\mathcal{U}^{\text{meas}\dagger}\, \mathcal{U}^\text{meas}\ket{\psi}_{S}\ket{\phi}_{E} = \braket{\psi}{\psi}_{S} \braket{\chi}{\phi}_{E}
	\end{equation}
	and therefore
	\begin{equation} \label{eq:check_orthogonality}
		\braket{\chi}{\phi}_{E} = \sum_{ij}{\braketOP{\psi}{\bbPi^\dagger_i\bbPi_j}{\psi}_{S} \braket{\mathcal{E}_{i\chi}}{\mathcal{E}_{j\phi}}}_{E}.
	\end{equation}
	
	Using the properties of orthonormal projective measurements \cref{eq:orthonormal_projective_ieasurement:hermitian,eq:orthonormal_projective_ieasurement:orthonormal} we get
	\begin{equation} \label{eq:conv_set_eq}
		\braket{\chi}{\phi}_{E} = \sum_i{p_i\alpha^i_{\chi,\phi}},
	\end{equation}
	where \(p_i= \braketOP{\psi}{\bbPi^\dagger_i\bbPi_i}{\psi}_{S}\) and \(\alpha^i_{\chi, \phi} = \braket{\mathcal{E}_{i\chi}}{\mathcal{E}_{i\phi}}_{E}\). Due to a property of projective measurement (\cref{eq:orthonormal_projective_ieasurement:sum}) we see that:
	\begin{equation}
		\sum_i{p_i} = \sum_i{\braketOP{\psi}{\bbPi_i}{\psi}}_{S} = \braketOP{\psi}{\Big(\sum_i\bbPi_i\Big)}{\psi}_{S} = \braket{\psi}{\psi}_{S} = 1.
	\end{equation}
	
	The numbers \(\{\alpha^i_{\chi,\phi} \text{, for all } i\}\) are distributed in the complex plane (\cref{fig:convex_sets}) and the expression \(\sum_i{p_i\alpha^i_{\chi,\phi}}\) is a convex combinations of these numbers. By varying/choosing \(\ket{\psi}_{E}\) we get different sets of number \(\{p_i\}\); and thereby the sum \(\sum_i{p_i\alpha^i_{\chi,\phi}}\) spans the entire convex set formed by conv\(\{\alpha^i_{\chi,\phi}\}\) as in \cref{fig:convex_sets}. For \cref{eq:conv_set_eq} to hold always --- for every convex combination of \(\{\alpha^i_{\chi,\phi}\}\) to equal the same point \(\braket{\chi}{\phi}_{E}\) --- the shaded area has to have zero measure and be at \(\braket{\chi}{\phi}_{E}\). This guarantees that the only solution is the trivial one: \(\alpha^i_{\chi,\phi} = \braket{\chi}{\phi}_{E}\) for all \(i\). Combining this result with \cref{stat:orthogonality}, we see that the inner products are preserved on the environmental states conditioned on the same outcome, or generally:
	\begin{equation}
		\braket{\mathcal{E}_{i\chi}}{\mathcal{E}_{j\phi}}_{E} = \delta_{ij}\braket{\chi}{\phi}_{E}.
	\end{equation}
	
	\begin{figure} [h!]
		\centering
		\begin{tikzpicture}
			\draw[->] (-1,0) -- (3,0) node [below] {\(\mathfrak{Re}\)};
			\draw[->] (0,-1) -- (0,3) node [right] {\(\mathfrak{Im}\)};
			
			\fill[pattern color=blue, pattern=north east lines] (1,-0.4) -- (-0.25,-0.5) -- (-0.5,0.75) -- (0,2) -- (1.5,1.75) -- (1.7,0.5) -- cycle;
			
			\node at (0.5,1) {{\color{red} \textbf{x}}};
			\node at (-0.5,0.75) {{\color{red} \textbf{x}}};
			\node at (1.5,1.75) {{\color{red} \textbf{x}}};
			\node at (0,2) {{\color{red} \textbf{x}}};
			\node at (-0.25,-0.5) {{\color{red} \textbf{x}}};
			\node at (1.7,0.5) {{\color{red} \textbf{x}}};
			\node at (0.5,0.5) {{\color{red} \textbf{x}}};
			\node at (1,-0.4) {{\color{red} \textbf{x}}};
		\end{tikzpicture}
		\caption{\label{fig:convex_sets} The points \(\{\alpha^i_{\chi,\phi}\}\) on the complex plane are marked with red crosses. The convex hull of these points corresponds to the blue hatched region. In case every combination of points in this convex hull equals a single complex number \(\braket{\chi}{\phi}_{E}\) (\cref{eq:conv_set_eq}), the hull has zero measure and trivially equals that complex number.}
	\end{figure}
	
	\subsection{Knill-Laflamme Structure and Quantum Error Correction}
	\label{sec:kl:qec}
	We now investigate the structure induced on \(\mathcal{H}_{E}\) due to the requirement that the post-measurement conditional environment states are distinct and orthogonal. These restrictions select a special subspace in \(\mathcal{H}_{E}\), which maps to a tensor factorization containing independent information about the pointer outcome and a memory (of the prior environment) state before measurement. We specifically explore connections between this subspace and quantum error correction, particularly the Knill-Laflamme conditions, which specify the types of errors that can be corrected. In our case, it specifies which unitary maps from \(\Phi\) yield distinguishable results.
	
	We define \(\bbPi_\Phi\) as the projector onto the \(\Phi\) subspace of ``good'' initial environment states: \(\Phi = \bbPi_\Phi\mathcal{H}_{E}\). As in \cref{sec:corr_res}, we denote the post-measurement subspace corresponding to outcome \(i\) as \(\Phi_i= \{\ket{\mathcal{E}_{i \phi}}_{E}\text{, for all }  \ket{\phi}_{E} \in \Phi\}\). As the inner product is preserved, the measurement unitary in \cref{stat:measurement} can be written in terms of the von-Neumann measurement model as \(\mathcal{U}^\text{meas} = \sum_i{\bbPi_i\otimes \mathcal{U}_i}\) (\cref{eq:cond_unitary}) where \(\mathcal{U}_i\) is an operation that unitarily takes vectors in \(\Phi\) to \(\Phi_i\). It follows from \cref{stat:orthogonality} that distinct outcomes should have distinguishable environmental states. That is, for \(\ket{\phi}_{E} \in \Phi\), we get
	\begin{equation}
		\label{eq:distinguish_cond}
		\braket{\mathcal{E}_{i\phi}}{\mathcal{E}_{j\phi}}_{E} = \delta_{ij} \implies  \braketOP{\phi}{\mathcal{U}^{\dagger}_{i} \mathcal{U}_{j}}{\phi}_{E} = \delta_{ij}.
	\end{equation}
	Thus, on the subspace \(\Phi\), the conditional unitaries obey
	\(\mathcal{U}_i^\dagger \mathcal{U}_j = \delta_{ij}\mathbbm{1}\), and we obtain 
	\begin{equation}
		\label{eq:KL}
		\bbPi_\Phi\mathcal{U}_i^\dagger \mathcal{U}_j\bbPi_\Phi = \delta_{ij} \bbPi_\Phi.
	\end{equation} 
	This is a special case of the \emph{Knill-Laflamme} (KL) conditions in quantum error correction \cite{PhysRevLett.84.2525, PhysRevA.55.900, kribs2006operatorquantumerrorcorrection}. The general KL condition \(\bbPi_{\mathcal{C}} \tilde{E}^{\dagger}_{a} \tilde{E}_{b} \bbPi_{\mathcal{C}} = C_{ab} \bbPi_{\mathcal{C}}\) specifies the set of errors \(\{\tilde{E}_a\}\) which are correctable on a code subspace \(\mathcal{C}\) where \(C_{ab}\) are elements of a Hermitian matrix depending only on the errors. In our case, the emergence of distinct, orthogonal measurement outcomes parallels the KL condition, ensuring that the error operators (the conditional unitaries \(\{\mathcal{U}_{i}\}\)) map the code subspace \(\Phi\) to mutually orthogonal subspaces.
	
	To see the structure the KL condition imposes on \(\mathcal{H}_{E}\), consider the \emph{projected evolution maps} \(\bar{E}_i\), one for each of the \(d\) possible measurement outcomes (\cref{eq:num_meas_out}),
	\begin{equation}
		\bar{E}_i:= \mathcal{U}_i\bbPi_{\Phi}, \quad i= 0,1,\dots,d-1.
	\end{equation}
	Let \(\dim \Phi = m\) so that the projector \(\bbPi_{\Phi}\) is a rank \(m\) projector. In terms of these projected maps, we express our KL condition of \cref{eq:KL} as
	\begin{equation}
		\bar{E}_i^{\dagger}\bar{E}_j =\delta_{ij} \bbPi_{\Phi}.
	\end{equation}
	Each of the \(\bar{E}_i\)s map the subspace \(\Phi\) into the distinct, orthogonal subspaces, \(\Phi_i= \mathcal{U}_i\bbPi_\Phi\mathcal{H}_{E}\) ensuring \(\Phi_i\perp \Phi_j\) for \(i\neq j\). As discussed above, these projected maps preserve the inner product \(\braket{\chi}{\phi}_{E}\) under evolution; each \(\bar{E}_i\) is an isometry from \(\Phi\) to \(\Phi_{i}\). In particular, \(\bar{E}_i^{\dagger}\bar{E}_i=\bbPi_\Phi\) is the projector on the initial subspace \(\Phi\), and \(\bar{E}_i\bar{E}_i^{\dagger}\) is the projector onto the final space \(\Phi_i\). Because of this, the dimension of each \(\Phi_i\) is the same as the dimension of \(\Phi\): \(\dim \Phi_i=\dim \Phi = m\) for all \(i\).
	
	Next, we consider the union of these disjoint subspaces, which we will call  \(\mathcal{H}_{KL}:=\bigoplus_{i=0}^{d-1}\Phi_i\). States in \(\mathcal{H}_{KL}\) follow the Knill-Laflamme distinguishability condition, \cref{eq:KL}. The dimension of \(\mathcal{H}_{KL}\) is \(md\) since there are \(d\) such orthogonal subspaces, each with dimension \(m\). We can now use this to induce a tensor structure on \(\mathcal{H}_{KL}\) which accounts for both prior information about the initial environment state and pointer information post-measurement. We choose an orthonormal basis for \(\Phi\) so that \(\Phi = \text{span} \{\ket{\mu}_{E}\}_{\mu=0}^{m-1}\). Each pair \(i,\mu\) labels a particular measurement outcome \(i\) with the initial environment basis state \(\ket{\mu}_{E} \in \Phi\). Let us define the state \(\ket{i,\mu}_{E}\) as \(\ket{\mu}_{E}\) mapped into \(\Phi_{i}\):
	\begin{equation}
		\ket{i,\mu}_{E} = \bar{E}_i\ket{\mu}_{E} \in \Phi_i.
	\end{equation}
	By the KL condition of \(\bar{E}_i^{\dagger}\bar{E}_j =\delta_{ij} \bbPi_{\Phi}\), the kets \(\ket{i,\mu}_{E}\) for all \(i,\mu\) form an orthonormal basis of \(\mathcal{H}_{KL}\):
	\begin{equation}
		\braket{i,\mu}{j,\nu}_{E} = \braketOP{\mu}{\bar{E}_i^{\dagger} \bar{E}_j}{\nu}_{E} = \delta_{ij}\braket{\mu}{\nu}_{E}  = \delta_{ij} \delta_{\mu\nu} .
	\end{equation}
	
	We introduce an abstract \(d\)--dimensional space \(\mathcal{H}_{P} \simeq \mathbb{C}^{d}\) (``P'' used to denote ``pointer'' which stores information about the corresponding measurement outcome) with orthonormal basis \(\mathcal{H}_{P}  = \text{span}\{\ket{i}_{P}\}_{i = 0}^{d-1}\). Similarly, we introduce an abstract \(m\) dimensional Hilbert space \(\mathcal{H}_{M} \simeq \mathbb{C}^{m}\) for the ``memory'' which keeps track of the information of the initial environment state. Following this, we define a unitary map \(W\),
	\begin{equation} \label{eq:unitary_tensor_embed} 
		W: \ket{i}_{P} \otimes\ket{\mu}_{M} \to \ket{i,\mu}_{E} \equiv \bar{E}_i\ket{\mu}_{E},
	\end{equation}
	which defines the isomorphism of \(\mathcal{H}_{KL}\) into a tensor product structure:
	\begin{equation}
		\mathcal{H}_{KL} \simeq \bigoplus_{i=0}^{d-1}\Phi_i\simeq \mathcal{H}_{P} \otimes \mathcal{H}_{M}.
	\end{equation}
	
	This is consistent with the no-deletion theorem, which states that no quantum information can be deleted during the measurement unitary. There are three possibilities for the relationship of the initial subspace \(\Phi\) with respect to \(\mathcal{H}_{KL}\):
	\begin{enumerate}[label=(\roman*)]
		\item \(\Phi\) and \(\mathcal{H}_{KL}\) could be disjoint,
		\item \(\Phi = \Phi_{i_\star}\) for a particular index \(i_\star\), or,
		\item there could be a partial overlap between \(\Phi\) and at least one of the \(\Phi_{i_\star}\)s.
	\end{enumerate}
	In every case, the environment Hilbert space decomposes as
	\begin{equation}
		\mathcal{H}_{{E}} \simeq  \mathcal{H}_{{KL}} \oplus \mathcal{H}_{\text{rest}}
	\end{equation}
	where \(\mathcal{H}_{\text{rest}}\) is taken as the environment subspace where measurement is ineffective. \(\mathcal{H}_{\text{rest}}\) could be the null set.
	
	In the language of operator algebra quantum error correction \cite{kribs2006operatorquantumerrorcorrection, PhysRevLett.100.030501, 10.5555/2012086.2012092}, this corresponds to the Wedderburn decomposition (see \cite{Wedderburn1934, bény2020algebraicapproachquantumtheory} or \cite[appendix A]{Harlow_2017}) of the environment’s Hilbert space. The \(d^{2}\) operators, \(F_{ij}:= \bar{E}_iE^\dagger_j\) generate the full matrix algebra on \(\mathbb{C}^{d}\); it is a matrix basis for \(\mathcal{L}(\mathbb{C}^{d})\) which we show generates an algebra on the pointer factor \(\mathcal{H}_{{P}}\). They are closed under Hermitian adjoint, \(F_{ij}^{\dagger} = F_{ji}\) and are orthogonal in the sense of \(F_{ij}F_{kl} = \delta_{jk}F_{il}\). These operators \(F_{ij}\) can be written in the tensor basis \(\ket{k}_{P}\otimes\ket{\mu}_{M}\) as a unitary change of basis through \(W^{\dagger} F_{ij} W\); \(W\) as defined in \cref{eq:unitary_tensor_embed}, is the embedding function of \(\mathcal{H}_{P}\otimes\mathcal{H}_{M}\) into \(\mathcal{H}_{E}\). To do so, we consider its action on a basis state \(\ket{k}_{P}\otimes\ket{\mu}_{M}\),
	\begin{equation}
		\big(W^{\dagger} F_{ij} W\big) \ket{k}_{P}\otimes\ket{\mu}_{M} = W^{\dagger} \bar{E}_i\bar{E}_j^{\dagger} E_{k}\ket{\mu}_{E}.
	\end{equation}
	Then using the KL condition \(\bar{E}_j^{\dagger}E_k =\delta_{jk} \bbPi_{\Phi}\) of \cref{eq:KL}, we get
	\begin{equation}
		\big(W^{\dagger} F_{ij} W\big) \ket{k}_{P}\otimes\ket{\mu}_{M}= \delta_{jk} W^{\dagger}\bar{E}_i \ket{\mu}_{E} =  \delta_{jk} W^{\dagger} \ket{i,\mu}_{E} .
	\end{equation}
	Now \(W^{\dagger}\) will map \(\ket{i,\mu}_{E}\) to the explicit tensor product, giving the final result as
	\begin{equation}
		\big(W^{\dagger} F_{ij} W\big)  \ket{k}_{P}\otimes\ket{\mu}_{M} = \delta_{jk} \ket{i}_{P}\otimes\ket{\mu}_{M}.
	\end{equation}
	Moreover, this tells us that \(F_{ij}\) acts as an identity on the memory factor and acts as a basis switch in the pointer space within the KL subspace:
	\begin{equation}
		W^{\dagger} F_{ij} W = \ketbra{i}{j}_{P} \otimes \mathbbm{1}_{M}.
	\end{equation}
	Notice that \(F_{ii} = \bar{E}_i\bar{E}_i^{\dagger}\) is a projector on \(\Phi_i\), and thus a projector on \(\mathcal{H}_{KL}\), further resolving the identity on the \(\mathcal{H}_{KL}\) subspace as \(\sum_{i=0}^{d-1}{F_{ii}} = \mathbbm{1}_{{KL}}\). It should be noted that the \(F_{ij}\) operators act on the abstract Hilbert space \(\mathcal{H}_{P} \otimes \mathcal{H}_{M}\). The projected map \(\bar{E}_i\) in general does \emph{not} need to be in this algebra, which, for example, will be the case when \(\Phi\) is disjoint from \(\mathcal{H}_{{KL}}\). Overall, we establish a connection between the redundant records of quantum Darwinism and objective classical reality, utilising structures in quantum error correction. We hope to explore this connection in further detail in future work.
	
	To understand the action of the unitary measurement procedure for initial states in \(\Phi\) in relation to this KL subspace, we can (somewhat artificially) decompose each projected unitary \({E}_{i} \equiv \mathcal{U}_{i}\bbPi_{\Phi}\)  acting on \(\ket{\phi}_{E} = \sum{\phi_\mu\ket{\mu}_{E}} \in \Phi\) as the composition of two operations, \(\mathcal{U}_{i}\ket{\phi}_{E} = {C}_{i_\star\to i} \circ {E}_{i_\star}\ket{\phi}_{E}\):
	\begin{enumerate}
		\item \emph{\uline{Readying:}} the operation \({E}_{i_\star}\) brings the initial environment \(\ket{\phi} \in \Phi\) to within the KL subspace, particularly to a ``ready'' state with a fiducial pointer value \(i_\star\). That is, \(W^\dagger {E}_{i_\star} \ket{\phi}_{E} = \ket{i_\star}_{P} \otimes \ket{\phi}_{M}\) for a memory state \(\ket{\phi}_{M} = \sum{\phi_\mu\ket{\mu}_{M}} \in \mathcal{H}_{{M}}\) which contains all information about the initial state (in accordance with no-deletion).
		\item \emph{\uline{Correlating:}} the operation \({C}_{i_\star\to i}\) creates a correlated record of the signal state \(i\) in the environment, \((W^\dagger{C}_{i_\star\to i} W)\ket{i_\star}_{P} \otimes \ket{\phi}_{M} = \ket{i}_{P} \otimes \ket{\phi}_{M}\) by mapping \(\Phi_{i_\star}\) to \(\Phi_{i}\). This latter step of creating a correlated record with the system can be performed entirely in the abstract \(\mathcal{H}_{P}\otimes\mathcal{H}_{M}\) subspace. 
	\end{enumerate}
	
	In our entirely arbitrary separation into readying and correlating, we have made an arbitrary choice of ``ready pointer \(i_\star\)''. We could have chose some other value for this ready pointer (say \(i'_\star\)) and a corresponding division into readying and correlating operators to achive the same results: \(\mathcal{U}_{i}\ket{\phi}_{E} = {C}_{i'_\star\to i} \circ {E}_{i'_\star}\ket{\phi}_{E}\). The readying operation \({E}_{i_\star}\) maps the special initial subspace \(\Phi\) to a ``ready'', well-defined pointer state (which is well suited for correlating with the system), and moves the initial information of the environment from \(\Phi\) to the memory factor \(\mathcal{H}_{M}\) of the KL subspace \(\mathcal{H}_{{KL}}\). The readying operation is akin to \emph{environmental correction} (as alluded to in the Stern-Gerlach example in \cref{sec:intro}), where the environment corrects its state to allow the creation of correlation with the system, and as a consequence, stores its prior information in the memory sector. We shall illustrate this in a concrete example in the next section. In the special case when \(\Phi \subset \mathcal{H}_{{KL}}\), that is, with \(\Phi = \Phi_{{k_\star}}\) for some \(k_\star\), we have our readying operation fully within the pointer algebra, \({E}_{i_\star} = \mathcal{U}_{i_\star} \bbPi_{\Phi_{k_\star}} = F_{i_\star k_\star}\). However, even in other cases, we can always use \({E}_{i_\star}\) to bring the subspace \(\Phi\) to the \(i_\star\)-ready-state-subspace \(\Phi_{i_\star}\) and then use the \(F_{ij}\) operators to correlate.
	
	Here, we have taken a top-down approach. That is, we look at the structure induced on \(\mathcal{H}_{{E}}\) (for both the initial subspace \(\Phi\) and post-measurement subspace \(\mathcal{H}_{{KL}}\)) by demanding that the final state of the signal and environment after measurement take the form of \cref{eq:quantum_ieasurement} with distinguishable outcomes. The above-described splitting of the measurement procedure into \emph{readying} and \emph{correlating} is artificial and serves to define the algebra of the operators abstractly, thereby placing limits on the initial state of the environment. It provides a coarse-grained perspective on the measurement procedure in the top-down tradition. This approach views measurement as a one-shot process, overlooking the details of the microdynamics that enable it to occur in a manner consistent with spacetime local interactions (due to special relativity) and quantum mechanics. On the other hand, in \cref{sec:meas_proc}, we will look at a concrete example of a unitary measurement procedure, and thereby take on a bottom-up approach instead --- constructing the measurement procedure from spacetime local interactions. We will define an explicit microdynamics and outline its connection with the KL algebra and the top-down approach taken in this section.
	
	\subsubsection{Generalized Pauli algebra:}
	We now provide a specific implementation of the action of the projected maps (vis-à-vis readying and correlating) by making contact with the Generalized Pauli algebra \cite{singh2020modelingpositionmomentumfinitedimensional,PhysRevA.57.127, Jagannathan:2010sb, SanthanamTekumalla1976, Jagannathan:1981ri}. By choosing a specific bijection between the set \(\{i\}\) (of size \(d\)) and the integers \(\{0,1,\dots,d-1\}\), we can define the generalized Pauli \(X\) operator on \(\mathcal{H}_{{KL}}\) as
	\begin{equation}
		X = \sum_{l=0}^{d-1} F_{l+1, l},
	\end{equation}
	with cyclic boundary conditions identifying \(l = d\) with \(l = 0\), thereby working in modulo-\(d\) arithmetic. The \(X\) operator acts as a generalized \emph{shift} operator on \(\mathcal{H}_{{KL}}\),
	\begin{equation}
		\big(W^\dagger X W\big) \ket{l}_{P}\otimes\ket{\mu}_{M} = \ket{l+1}_{P}\otimes\ket{\mu}_{M}.
	\end{equation}
	
	One can similarly define a generalized \emph{clock} operator, \(Z\) conjugate to the \(X\) as,
	\begin{equation}
		Z = \sum_{l=0 }^{d-1}\omega^l F_{ll},
	\end{equation}
	where \(\omega = e^{2\pi i/d}\) is the primitive root of unity. The action of \(Z\) on the KL subspace is
	\begin{equation}
		\big(W^\dagger Z W\big) \ket{l}_{P}\otimes\ket{\mu}_{M} = \omega^l \ket{l}_{P}\otimes\ket{\mu}_{M}.
	\end{equation}
	The algebra is closed under \(X^{d} = Z^{d} = \mathbbm{1}_{{KL}}\) and satisfies the Weyl commutation relation \cite{weyl1950theory}: \(ZX=\omega XZ\). We emphasize that the identification of this generalized Pauli algebra is non-unique. A permutation of the labels associated with the integers \(\{0,\dots,d-1\}\) gives a different algebra with its own version of the \(X\) operator. Any such permutation produces a valid algebra that closes as \(ZX=\omega XZ\); these different bijections correspond to unitarily equivalent versions of \(X\) and \(Z\).
	
	With the \(X\) operator above we can rewrite \cref{eq:cond_unitary} with the operator:
	\begin{equation} \label{eq:quantum_ieasurement_pauli_x}
		\mathcal{U}^\text{meas} = \left(\sum_i{\bbPi_i\otimes X^{i - i_\star}}\right) \circ\Big(\mathbbm{1}_{S}\otimes \bar{E}_{i_\star}\Big) = \sum_i{\bbPi_i\otimes \Big( X^{i - i_\star}}  \bar{E}_{i_\star}\Big),
	\end{equation}
	where we can now identify the readying operation \(\bar{E}_{i_\star}\) which brings the environment state into a ready state within the KL subspace, and the correlating operation which we write in terms of the generalized Pauli operator \(C_{i_\star \to i} = X^{i - i_\star}\). We emphasize again that \(X\) only acts on the KL subspace of the environment, and it is a coarse-grained operator that does not account for all the details regarding how operations are defined in spacetime. In fact, as \(F_{ij}\) are only defined within the KL subspace and since \(\bar{E}_i\) in general does not lie in their span, the readying operation mapping to a ready state \(i_\star\) is essential. With respect to quantum error correction, the readying part corresponds to error correction that brings the state of the environment to a well-defined one within the KL subspace. The correlating part is what imprints the state of the signal onto the environment. In the following subsection, we examine the extension needed when multiple pointers redundantly record the signal state.
	
	\subsection{Redundant Records from Correlated Multi-partite Environments} \label{sec:kl:multi_env}
	A crucial feature of emergent classicality is the presence of multiple, redundant records of the signal state. In the language of the previous subsection, this means that the environment does not consist of a single pointer, but rather of several distinct fragments, each capable of recording the signal outcome independently \cite{PhysRevLett.105.020404, Riedel_2011, PhysRevA.80.042111}. We now extend the structure developed in the single-pointer case to a multi-partite environment and examine how redundancy constrains the dimensionality of the Hilbert space and the special initial subspace \(\Phi\). We consider the environment to be composed of \(N\) fragments, each capable of forming an independent record,
	\begin{equation}
		\mathcal{H}_{{E}} \simeq \bigotimes_{\alpha=1}^{N} \mathcal{H}_{{E}_\alpha}.
	\end{equation}
	We emphasize that this decomposition is a functional decomposition into informational subsystems that participate in recording the signal outcome. What plays the role of a ``pointer'' or a ``memory'' (of the prior environment) degree of freedom depends on the algebraic identification within the environment’s Hilbert space --- the same physical degrees of freedom may be partitioned differently depending on the measurement interaction (see below for a concrete example).
	
	Each fragment \(\mathcal{H}_{E_\alpha}\) undergoes the same measurement interaction with the signal as in the single-fragment case. The KL structure derived previously in \cref{sec:kl:qec} induces a similar decomposition within each fragment, \(\mathcal{H}_{E_\alpha} \simeq \mathcal{H}_{{KL}_\alpha} \oplus \mathcal{H}_{\text{rest}_\alpha} \simeq \left(\mathcal{H}_{P_\alpha} \otimes \mathcal{H}_{M_\alpha}\right) \oplus \mathcal{H}_{\text{rest}_\alpha}\). Here \(\mathcal{H}_{P_\alpha}\) denotes the pointer factor \(\alpha\) within the KL subspace responsible for storing the measurement record. \(\mathcal{H}_{M_\alpha}\) refers to the memory factor of the corresponding fragment, which might carry information about the initial state of that fragment; it could very well be that the fragment is already in a ready state, and the memory factor is absent. The subspace \(\mathcal{H}_{\text{rest}_\alpha}\) represents any degrees of freedom that do not participate in the high fidelity measurement as described. Taking the tensor product of all fragments, the global KL subspace of the environment becomes
	\begin{equation}
		\mathcal{H}_{{KL}} \simeq
		\bigotimes_{\alpha=1}^{N} \mathcal{H}_{{KL}_\alpha}
		\simeq \left(\bigotimes_{\alpha=1}^{N} \mathcal{H}_{P_\alpha}\right) \otimes \mathcal{H}_M,
		\label{eq:HKL-multipartite}
	\end{equation}
	where \(\mathcal{H}_M \equiv \bigotimes_\alpha \mathcal{H}_{M_\alpha}\) is the combined memory sector. The total environment factorizes as
	\begin{equation}
		\mathcal{H}_E \simeq  \mathcal{H}_{{KL}} \oplus \mathcal{H}_{\text{rest}}\:,
	\end{equation}
	where the remainder \(\mathcal{H}_{\text{rest}}\) collects all combinations of fragment subspaces that include at least one \(\mathcal{H}_{\text{rest}_\alpha}\), that is, any term that fails to participate in the distinguishable measurement process.
	
	In the single-fragment case, the correlating action of the measurement was implemented by the generalized Pauli shift operator \(X\), acting only on the pointer sector. In the multi-partite case, the system interacts with multiple pointer fragments simultaneously. We denote by \(K \le N\) the number of fragments that successfully record the signal outcome. The correlating operator on the environment now takes the form of a tensor product over pointers. While recognizing that the ready pointer label \(i_\star\) need not all align with each other, it is generally a multi-indexed operator,
	\begin{equation}
		X^{i - \{i_{\alpha\star}\}} = \bigotimes_{\alpha=1}^{K}{X_\alpha^{i-i_{\alpha\star}}}.
	\end{equation}
	The above correlating operator has different pointer ready states \(i_{\alpha\star}\) for its different fragments \(\alpha\). They pair with correspondingly different readying operators \(\bar{E}_{i_{\alpha\star}}\) to enable a consistent measurement procedure. We demand a single, coarse-grained \(X\) operator to act on all the pointer fragments which would correspond to the same ready state \(i_{\alpha\star} = i_\star\) for all pointers \(\alpha\). We, thereby, obtain a single correlating operator \(X^{i-i_\star}\). This alignment ensures that the pointer operators act coherently across all fragments, producing consistent records of the signal pointer state. This consistency requirement parallels the condition for SBSs in \cref{eq:sbs}: distinct environmental fragments must redundantly encode the same outcome \(i\), ensuring objectivity across observers. As the division into readying and correlating is artificial in any case, there is no loss of generality here --- we are transferring all environmental correction duties to the readying operators. Note that the readying and correlating operations do not commute in general. In the next section, we develop a specific microdynamics and establish a connection between the top-down approach presented here and the bottom-up approach we will develop there.
	
	Applying our potentially complicated readying operator \(\bar{E}_{i_\star}\) onto the initial state \(\ket{\phi}\in \Phi\) of the environment, we reach a ready state (notice the transfer of initial environment information to \(\mathcal{H}_{M}\))
	\begin{equation}
		\bar{E}_{i_\star}\ket{\phi} = \ket{i_\star}_{P_{1}} \ket{i_\star}_{P_{2}} \cdots \ket{i_\star}_{P_N} \otimes \ket{\phi}_{M} \in \Phi_{i_\star}.
	\end{equation}
	We get several pointers ready to create a record of a quantum measurement, and a memory sector that contains information about the initial state of the environment. We now analyse the dimensions of these spaces. The ready sector \(\Phi_{i_\star}\) has the same dimensions as the memory factor \(\Phi\). Each factor \(\mathcal{H}_{P_\alpha}\) has to have a dimension at least \(d\) to accommodate recording of the measurement outcome. There are at most \(N\) pointers \(P_\alpha\) available in the environment, of which \(K\) make redundant records. The memory factor has dimension of the initial state \(\dim{\Phi}\) and we therefore get, as \(\mathcal{H}_{E} \supseteq \mathcal{H}_{KL}\), 
	\begin{equation}
		\mathcal{H}_{E} \supseteq \left(\bigotimes_{\alpha=1}^{K}\mathcal{H}_{P_\alpha}\right) \otimes \mathcal{H}_{M}.
	\end{equation}
	It follows that
	\begin{equation}
		d^K\dim{\Phi} \le \left(\prod_\alpha\dim{\mathcal{H}_{P_\alpha}}\right)\dim{\mathcal{H}_{M}} \le\dim{\mathcal{H}_{E}}
	\end{equation}
	and thus:
	\begin{equation} \label{eq:dimension_inequality}
		K\ln(d) + \ln(\dim\Phi) \le \ln(\dim\mathcal{H}_{E}).
	\end{equation}
	We can express this inequality in an alternative way, offering an interpretation of the capacity in the environment to create redundant records,
	\begin{equation} \label{eq:dimension_inequality2}
		K \le \log_{d}(\dim\mathcal{H}_{E}) - \log_{d}(\dim\Phi).
	\end{equation}
	The number of redundant records that the environment can support is upper bounded by the residual capacity of the environment to fit \(d\) pointer states outcomes after accounting for the space needed by \(\Phi\) to store the information about the initial state of the environment prior to measurement in accordance with no-deletion.
	
	Seen differently, one can interpret the inequality \cref{eq:dimension_inequality2} as a constraint on the type of environment states allowed, specifically the need for \emph{correlation} among different environmental fragments. In particular, \(\log_{d}(\dim\Phi) \le \log_{d}(\dim\mathcal{H}_{E}) - K\) which implies that for more redundant records \(K\), the dimension of the initial environment subspace \(\Phi\) will be lower. Thus, for the state of \(N\) fragments to start in a lower-dimensional subspace implies that information must be shared between fragments, hence implying \emph{correlation} among them. The more redundant the measurement outcomes are, the greater the correlation (and smaller \(\dim \Phi\)) required in the initial environment.
	
	We consider a few concrete examples involving this dimensionality bound to understand how correlation is a finite resource in the environment consumed every time a measurement is performed. For concreteness and simplicity, we take the dimension of the environment to be of the form \(\dim{\mathcal{H}_{E}} = d^N\), that is, the environment is isomorphic to the Hilbert space of \(N\) qudits. Relevant subspaces of this \(d^N\)-dimensional space correspond to pointer factors and a memory, as we will now see. \Cref{eq:dimension_inequality} now takes the form:
	\begin{equation}
		K\ln(d) \le N\ln(d) - \ln(\dim\Phi).
	\end{equation}
	
	Let us first consider \(N=1\) so that there is only one environment qudit. In this case, if the initial environment state is in a ready state, \(\ket{\phi} = \ket{i_\star}_{E}\), the state space is one dimensional (\(\dim\Phi = 1\)) and we get \(K\ln(d) = N\ln(d) - \ln(1) \implies K \le 1\). Thus, one record can be formed. This is the situation usually discussed in decoherence texts, where the environment is already in a ready state. In the Stern-Gerlach example, it corresponds to the environment starting in the well-defined state \(\ket{\uparrow}_{A}\) leading to \cref{eq:stern_gerlach_correct,eq:stern_gerlach_correct_density} --- measurement takes place successfully. On the other hand, the environment could start in an arbitrary state, \(\ket{\phi} = \sum_l{\phi_l\ket{l}_{E}}\). In this case, \(\dim\Phi =\dim\mathcal{H}_{{E}} = d\) and we get \(K\ln(d) = N\ln(d) - \ln(d) \implies K \le 0\). No records can be formed because there is no ``space'' for the initial information about the environment to be dumped. In the Stern-Gerlach example of \cref{sec:intro}, this corresponds to the environment starting in the arbitrary state \(\ket{\phi}_{A}\) leading to \cref{eq:stern_gerlach_dont_work,eq:stern_gerlach_dont_work-density} --- measurement does not work. This confirms the results of \cref{th:no-go} and is a reflection of the no-deletion theorem.
	
	Next, we look at multi-partite environments where redundant records can be created. We now consider the environment made up of \(N=3\) qudits (labelled \(Q_{1}, Q_{2}, Q_{3}\)), \(\mathcal{H}_{E} \simeq \mathcal{H}_{Q_{1}} \otimes \mathcal{H}_{Q_{2}} \otimes \mathcal{H}_{Q_{3}} \simeq (\mathbb{C}^d)^{\otimes3}\) and we choose, for illustration, the ``good'' initial subspace to be a \(d\)-dimensional subspace \(\dim\Phi=d\) (same as the dimension of the environment qudits). The following are a few different (non-exhaustive) examples of such a state \(\ket{\phi} \in \Phi\), each corresponding to a different subspace (\(\Phi\)), albeit with the same dimension \(\dim\Phi=d\):
	\begin{subequations} \label{eq:different_initial_envs}
		\begin{align}
			\ket{\phi}_{E} &= \ket{i_\star}_{Q_1}\sum_l{\phi_l\ket{l}_{Q_2}\ket{l}_{Q_3}}, \\
			\ket{\phi}_{E} &= \sum_l{\phi_l\ket{l}_{Q_1}\ket{l}_{Q_2}\ket{l}_{Q_3}}, \\
			\ket{\phi}_{E} &= \sum_l{\phi_l\ket{l+a}_{Q_1}\ket{l+b}_{Q_2}\ket{l}_{Q_3}}, \quad \text{for some fixed a and b}. \label{eq:different_initial_envs:crazy_case}
		\end{align}
	\end{subequations}
	After suitable readying operations, each of the above systems would be in the state
	\begin{equation}
		\bar{E}_{i_\star}\ket{\phi} = \ket{i_\star}_{Q_1}\ket{i_\star}_{Q_2}\otimes \left(\sum_l{\phi_l\ket{l}_{Q_3}}\right),
	\end{equation}
	allowing us to identify two of the qudits (\(Q_{1}\) and \( Q _ {2}\) here) as pointer registers (\(P_{1}\) and \(P_{2}\) here, respectively) available for correlating with the signal. In these cases, \(\mathcal{H}_{Q_{3}}\) acts as the memory \(\mathcal{H}_{M}\) and stores the initial information \(\{\phi_l\}\) of the environment\footnote{Since each of the qudits is \(d\)-dimensional and taken to be just as large as the number of system pointer outcomes, and since \(\dim\Phi\) is also chosen to be \(d\), we exhaust the environment state space giving us \(\mathcal{H}_\text{rest}\) to be the null space.}. While each of the representative states of \cref{eq:different_initial_envs} belong to different subspaces \(\Phi\) and the ``readying'' operation required to align them may differ, they have the same potential to create redundant records: \(K\ln(d) = N\ln(d) - \ln(d) \implies K \le 2\).
	
	To create these records, the initial environment had to be in a state that was correlated across multiple environment qudits as illustrated in \cref{eq:different_initial_envs}. On the other hand, a higher \(\dim \Phi\), for instance \(\dim \Phi = d^{2}\), would imply a less correlated state among the different qudits, for example,  \(\ket{\phi}_{E} = \sum_l{\phi_{lk}\ket{l}_{Q_1}\ket{l}_{Q_2}\ket{k}_{Q_3}}\). The readying operation could give rise to the ready state \(\bar{E}_{i_\star}\ket{\phi} = \ket{i_\star}_{Q_1}\otimes \left(\sum_l{\phi_{kl}\ket{l}_{Q_2}\ket{k}_{Q_3}}\right)\) which creates only one redundant record in accordance with \cref{eq:dimension_inequality}. With respect to the Stern-Gerlach example, this is the situation leading to the result in \cref{eq:stern_gerlach_make_work}.
	
	While the memory factor need not neatly fit into powers of \(d\), the inequality in \cref{eq:dimension_inequality} holds generally, indicating that more redundant records demand higher correlated initial environments. For the remainder of the article, we will focus on a memory dimension \(\dim\Phi = d\), as this is one of the simplest yet non-trivial cases that produces a large number of redundant records while still allowing the initial environment to carry some information. In this case, for a generic \(d^N\) dimensional environment, we get: \(K \le N-1\). For the following section, where we develop a concrete example, we stick to an initial environment of the form: \(\ket{\phi}_{E} = \sum_l{\phi_l\ket{l}_{E_1}\dots\ket{l}_{{E}_N}}\). It has the potential to create \(N-1\) records. In summary, multi-partite redundancy arises naturally from the KL structure when multiple pointer subsystems share a common outcome label. \Cref{eq:dimension_inequality} quantifies the resource tradeoff: for a finite-dimensional environment, the number of redundant records that can be created is bounded by the logarithmic difference between the total environmental dimension and the size of the special subspace \(\Phi\). This bound encapsulates the idea that \emph{correlation is a finite resource} --- each redundant measurement consumes part of the environment’s capacity to sustain consistent, objective records of the signal state.
	
	\section{Qudit Measurement Procedure} \label{sec:meas_proc}
	In this section, we consider a concrete instantiation of the measurement procedure by giving explicit unitaries and quantum systems on which they act. This construction provides a concrete, bottom-up model of a unitary measurement process acting on a finite-dimensional system and its correlated environment --- one that specifies the underlying microdynamics. The environment consists of \(N\) physical qudits that interact locally with the signal. This concrete realization embeds the Knill-Laflamme (KL) subspace, derived in \cref{sec:kl}, within the entire environment Hilbert space, illustrating how the readying and correlating operations arise from spacetime-local unitaries. The model also clarifies how correlation, as a finite resource, is transferred from the environment to the signal-observer network during measurement.
	
	We consider the signal Hilbert space to be spanned by its pointer basis, \(\mathcal{H}_{S} = \text{span}\{\ket{i}_{S}\}_{i=0}^{d-1}\), giving \(d\) one-dimensional projectors \(\bbPi_i = \ketbra{i}{i}_{S}\). The environment is modeled as a set of \(N\) \emph{physical} qudits, \(\mathcal{H}_{{E}} \simeq (\mathbb{C}^{d})^{\otimes N}\), (parts of which) actively participate in the measurement process. We remind the reader that the KL subspace embedding occurs within these physical qudits in a non-trivial manner, capturing the informational subsystems that participate in recording the signal outcome. We take the environment to begin in a correlated initial state belonging to the special \(d\)-dimensional subspace \(\Phi \subset \mathcal{H}_E\),
	\begin{equation}
		\ket{\phi}_E = \sum_{l=0}^{d-1} \phi_l\,\ket{l}{E_1}\ket{l}_{E_2}\cdots\ket{l}_{{E}_N} \in \Phi \subset \mathcal{H}_{E}.
		\label{eq:correlated-environment}
	\end{equation}
	This state, which we refer to as a \emph{correlated environment}, contains non-local correlations among its qudits such that each subsystem agrees on a common label \(l\) on every branch of the superposition. The coefficients \(\{\phi_l\}\) collectively encode the environment’s initial state. The correlated environment may be viewed as a \emph{correlation donor} --- a cohesive and internally ordered system that actively participates in the measurement process by supplying correlation to the signal, as opposed to a random or chaotic environment that merely induces decoherence. For example, in the Stern-Gerlach experiment, the magnet’s macroscopic field arises from a large ensemble of internally correlated (aligned) spins, providing a physical realization of such a correlated environment. \Cref{eq:correlated-environment} therefore describes an environment whose subsystems are perfectly correlated in the pointer basis, establishing the structure necessary for transferring correlation to the signal during measurement. 
	
	\subsubsection{Imprint operator:}
	The measurement procedure described here is a particular implementation of the microdynamics; we make the connection to the coarse-grained (top-down) dynamics of the previous section later. The spacetime local interactions we use make use of an \emph{imprint operator,} \(\mathcal{I}\), which acts on a pair of qudits, defined as
	\begin{equation} \label{eq:imprint}
		\mathcal{I}_{{a} \to {b}} \ket{i}_{a} \ket{j}_{b} = \ket{i}_{a} \ket{j+i}_{b},
	\end{equation}
	where \(j+i\) is addition modulo base \(d\). The imprint operation is basis dependent as \(j+i\) is defined with respect to the specific labels \(\{i\}\) associated with the basis \(\{\ket{i}\}\) and would look different in another basis. We also use the inverse of the imprint operation for our measurement procedure, defined as
	\begin{equation} \label{eq:imprint-inverse}
		\mathcal{I}^{-1}_{{a} \to {b}} \ket{i}_{a} \ket{j}_{b} = \ket{i}_{a} \ket{j-i}_{b}.
	\end{equation}
	
	These operators are local in the sense that they only involve contact between pairs of systems. While the top-down approach in \cref{sec:corr_res} does not need to concern itself with details of spacetime locality, the bottom-up approach in this section must adhere to the rules of spacetime locality of physical interactions as we do here\footnote{The imprint operation can also be represented by a Hamiltonian. The article \cite{Zurek_2000} writes the Hamiltonian as \(H_{{a} \to {b}} = \sum_{ij}{ij \ketbra{i}{i}_{a} \otimes \ketbra{j^\star}{j^\star}_{b}}\) where \(\mathcal{I}_{{a} \to {b}} = {e}^{-it_{int}H_{{a} \to {b}}}\) and \(t_{int}\) is the interaction time of the Hamiltonian. Moreover, \(\ket{j^\star} = \sum_l{{e}^{i 2 \pi nl/d}\ket{l}}\) is the quantum Fourier transform of \(\{\ket{l}\}\). If the interaction time is \(t_{int}=2\pi/d\), the result is the imprint operation of \cref{eq:imprint}. The inverse imprint can be generated by the Hamiltonian \(-H_{{a} \to {b}}\) acting for the same interaction time.}.
	
	\subsubsection{The measurement procedure:}
	Let the correlated environment start at the state in \cref{eq:correlated-environment} and the signal in state \(\ket{\psi}_{S} = \sum_i\psi_i\ket{i}_{S}\), as is always describable in the pointer basis. The overall state of the signal and the correlated environment is
	\begin{equation}
		\ket{\psi}_{S} \ket{\phi}_{E} = \sum_{il}{\psi_i\phi_l\ket{i}_{S}\ket{l}_{E_1}\ket{l}_{E_2}\dots\ket{l}_{{E}_N}}.
	\end{equation}
	
\begin{figure} [h!]
		\centering
		\includegraphics[width=\textwidth]{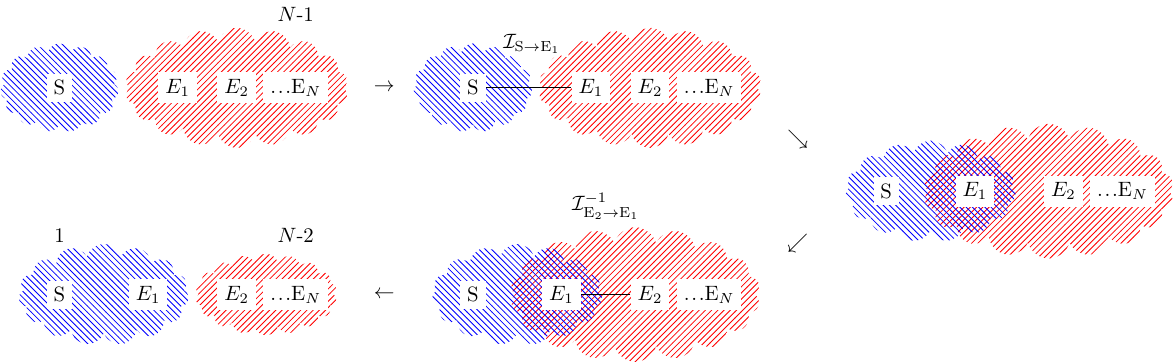}
		\caption{\label{fig:measurement-procedure} The process of entangling the signal and a subset of the environment to achieve measurement. Clouds indicate existing correlations, and lines between subsystems indicate local interactions. A correlation measure of \(N-1\) on the correlated environment distributes itself into a measure of \(N-2\) on the correlated environment and \(1\) on the signal observer network.}	
	\end{figure}

	The measurement procedure works as follows:
	\begin{enumerate}
		\item the signal imprints itself onto one qudit (say \(E_1\)) of the correlated environment (\(\mathcal{I}_{{S} \to {E_1}}\)),
		\item a neighbouring qudit of the correlated environment (\(E_2\)) performs environmental correction on the imprinted qudit \(E_1\) through an inverse imprint.
	\end{enumerate}
	
	\begin{equation} \label{eq:begin-measurement-procedure}
		\begin{alignedat}{2}
			\ket{\psi}_{S} \ket{\phi}_{E} &= \sum\psi_i\phi_l&&\ket{i}_{S}\ket{l}_{E_1}\ket{l}_{E_2}\dots\ket{l}_{{E}_N} \\
			&&&\Big\downarrow \mathcal{I}_{{S} \to {E_1}} \\
			&&&\ket{i}_{S}\ket{l+i}_{E_1}\ket{l}_{E_2}\dots\ket{l}_{{E}_N} \\
			&&&\Big\downarrow\mathcal{I}^{-1}_{{E_2} \to {E_1}} \\
			&&&\ket{i}_{S}\ket{l+i-l}_{E_1}\ket{l}_{E_2}\dots\ket{l}_{{E}_N} \\
			&\hookrightarrow \Big(\sum\psi_i&&\ket{i}_{S}\ket{i}_{E_1}\Big) \Big(\sum\phi_l\ket{l}_{E_2}\dots\ket{l}_{{E}_N}\Big)
		\end{alignedat}
	\end{equation}
	A visual summary is provided in \cref{fig:measurement-procedure}.
	
	If the above procedure continues with \(\mathcal{I}^{-1}_{{E_{\alpha+2}} \to {E_{\alpha+1}}} \circ \mathcal{I}_{{E_\alpha} \to {E_{\alpha+1}}}\) with \(\alpha\) taking on values \(1\), \(2\), and so on until \(K-1\), the resulting state is
	\begin{equation} \label{eq:correlated-env-state-after-k-measurements}
		\left(\sum_i\psi_i\ket{i}_{S}\ket{i}_{E_1}\dots\ket{i}_{{E}_K}\right) \left(\sum_l\phi_l\ket{l}_{E_{K+1}}\dots\ket{l}_{{E}_N}\right);
	\end{equation}
	the information about the signal measurement being (\(K\)-fold) redundantly recorded. This redundant recording indicates the generation of an objective classical reality about the measurement outcome.
	
	We call the state \(\sum_i\psi_i\ket{i}_{S}\ket{i}_{E_1}\dots\ket{i}_{{E}_K}\) an \emph{observer network state}. They are supposed to be (a particular instantiation of) the pure state equivalent of SBSs. In this particular instantiation, the redundant encoding of information about the signal measurement is manifest; however, this redundancy is a generic feature of quantum measurement once one accepts objective classical reality. Objective classical reality plays a two-fold role in the quantum measurement procedure it: quantifies the number of redundant records that are formed by a quantum measurement, \(K\) for the observer network state above, and, bounds the measurement procedure by placing constraints on the environment, which serves as a reservoir of correlation for the formation of correlated records, as in the special initial state of the environment in \cref{eq:correlated-environment}. \emph{Redundancy is both, resource and result}. It is essential to note that the role of the environment as a reservoir of correlation is specific to the pointer basis used for correlation. In the discussion, \cref{sec:discussion}, we expand on what happens when the correlation basis is slightly perturbed from the pointer basis of the environment: bounds are placed on the preferred basis problem.
	
	We have considered a particular microdynamics in this section. In this case, the imprint operation (most easily interpreted as the correlating operation) takes place before the inverse imprint (most easily interpreted as the readying operation, as it performs the environmental correction), in contrast to the previous section. While readying and correlating do not commute in general, the particular protocol we present here is consistent with previous developments. This works because the readying and correlating operations are one-shot operations on the entire Hilbert space of signal and environment. In contrast, the imprint operations are local, acting only between neighbours. We have decomposed the overall actions of readying and correlating by taking apart all the parts that can commute and recomposed them as a set of local operations. We use this specific setup to ensure consistency with the numerical results presented in the next section. We now demonstrate that the two approaches are equivalent.
	
	\subsubsection{Connecting the top-down and bottom-up perspectives:}
	We have covered the measurement process from both angles:
	\begin{enumerate}[label=(\roman*)]
		\item top-down --- due to restrictions of dimensionality, there is only a limited ability to create records, \cref{eq:quantum_ieasurement_pauli_x}, and,
		\item bottom-up --- a particular instantiation of a measurement procedure, as in this section, explains how the environment transfers correlation to a signal observer network using local interactions, \cref{eq:begin-measurement-procedure,eq:correlated-env-state-after-k-measurements}.
	\end{enumerate}
	Ultimately, the two perspectives must match when compared. The microdynamics explained in this section lay out in detail how correlation is transferred. Looking at the measurement procedure from a coarse-grained perspective, we find (with ready state \(i_\star=0\)): a readying operator
	\begin{equation}
		\bar{E}_0 = \sum_l{\left(\bigotimes_{\alpha=1}^{K}\ketbra{0}{l}_{E_\alpha}\right)\otimes\left(\bigotimes_{\alpha=K+1}^{N}\ketbra{l}{l}_{E_\alpha}\right)}
	\end{equation}
	and a correlating operator
	\begin{equation}
		\sum_l{ \ketbra{l}{l}_{S}\otimes X^{l-0} \otimes \left(\bigotimes_{\alpha=K+1}^{N}\mathbbm{1}_{E_\alpha}\right)}
	\end{equation}
	with \(X = \sum_k{\left(\bigotimes_{\alpha=1}^{K}\ketbra{k+1}{k}_{E_\alpha}\right)}\).
	
	The environment starts in the state \(\sum_l \phi_l \ket{l}_{E_1}\ket{l}_{E_2}\dots\ket{l}_{{E}_N}\) (\cref{eq:correlated-environment}) which, after the readying operation, becomes (in one shot)
	\begin{equation}
		\ket{0}_{E_1}\dots\ket{0}_{{E}_K}\sum_l \phi_l \ket{l}_{E_{K+1}}\dots\ket{l}_{{E}_N}.
	\end{equation}
	It has ready pointers \(E_1\) to \(E_K\) in state \(\ket{0}\) which then, along with signal \(\ket{\psi}_{S} = \sum_i\psi_i\ket{i}_{S}\) undergoes the overall correlating operation X to become
	\begin{equation}
		\left(\sum_i\psi_i\ket{i}_{S}\ket{i}_{E_1}\dots\ket{i}_{{E}_K}\right) \left(\sum_l\phi_l\ket{l}_{E_{K+1}}\dots\ket{l}_{{E}_N}\right).
	\end{equation}
	Thus, the coarse-grained dynamics of the previous section hide the details but generate the algebra. The bottom-up approach of this section demonstrates that such an algebra can be realised through local interactions.
	
	\subsubsection{Correlation capacity:}
	We term the number of redundant records the environment can create as its \emph{correlation capacity} --- a finite measure. Using \cref{eq:dimension_inequality2} we see that the number of records that can be created is correlation capacity \(= \log_d(\dim\mathcal{H_{E}}) - \log_d(\dim\Phi)\). In the particular case of this section, we get \(\log_d(d^N) - \log_d(d) = N-1\). This agrees well with our definition above of correlation capacity as the number of redundant records that can be generated --- if the procedure in \cref{eq:begin-measurement-procedure,eq:correlated-env-state-after-k-measurements} was performed up to \(K=N-1\), the environment would be exhausted of its correlation resource.
	
	\subsubsection{Measurement by an observer:}
	A crucial (anthropocentric) part of the environment above is an observer that records the measurement outcome. It is a subsystem of the environment and, therefore, taken to be correlated with it. The observer possesses a classical reality with respect to an environment. An individual observer could have started at an arbitrary initial state in its distant past, but is assumed to have become a part of the observer network state of the environment at some point through a quantum process similar to the measurement procedure described here. Let us consider the state we reach at the end of the measurement procedure, \cref{eq:correlated-env-state-after-k-measurements}; it is not a pre-measurement state but rather a fully decohered state. It is assumed to have interacted with the environment enough that a notion of reality can be consistently applied to its outcomes. As a result, the paradoxes in article \cite{PhysRevLett.126.130402} do not arise for this state. In principle, every state can be considered to be a reversible ``pre-measurement'' state, given a sufficiently powerful laboratory; the issue is one of practicality. The here constructed measurement outcome state in \cref{eq:correlated-env-state-after-k-measurements} is assumed to have interacted with enough other systems that it is not practically reversible.
	
	\section{Numerical Results} \label{sec:num_res}
	In \cref{sec:kl}, we analyzed the algebraic constraints that objective classical reality imposes on the environment. In particular, we showed that the formation of robust and redundant records forces the initial environment to lie within a special correlated subspace \(\Phi\), and that this subspace is mapped, under measurement dynamics, into mutually orthogonal conditional sectors satisfying the Knill-Laflamme condition of \cref{eq:KL}. \Cref{sec:meas_proc} then demonstrated, through an explicit spacetime local model, how this structure enables a correlated environment to transfer its internal correlation to the signal observer network and thereby generate redundant classical records. Up to this point, our analysis has focused on two extremes: an abstract characterization of the admissible subspaces \(\Phi\) and a highly structured example of a perfectly correlated environment, as shown in \cref{eq:correlated-environment}. What remains is an \emph{operational} question: given a generic quantum state of the environment, to what extent is it predisposed to form redundant, well-aligned records of the signal? More concretely:
	\begin{itemize}
		\item How should one quantify the ``degree of agreement’’ among environmental fragments?
		\item How does this correlation measure behave under a concrete measurement interaction?
		\item Can we study a wider sampling of initial environmental states and understand when record formation succeeds or fails?
	\end{itemize}
	
	We now illustrate this with a numerical study. Our numerical study is inspired by the article \cite{riedel-zurek}, which examined how classical information spreads across many fragments of an initially uncorrelated environment in the setting of quantum Darwinism. In contrast, our focus here is to explore how \emph{initial correlations} (or the absence thereof) influence the production of redundant records in the unitary measurement model developed earlier. To address these questions, we perform numerical simulations of a signal qudit interacting with \(N\) environmental qudits through pairwise local imprint operations as introduced in \cref{sec:meas_proc}.  We consider a global Hilbert space composed of \(N+1\) qudits of local dimension \(d\), where the subsystem labelled \(\alpha = 0\) plays the role of the signal \(S\) and the remaining \(\alpha = 1,2,\ldots,N\) constitute the environment \({E_\alpha}\). We use a qudit spin-chain for the environment to indicate pairwise locality: qudits \(N\) and 2 are pairwise local with qudit 1, qudits 3 and 1 are pairwise local with qudit 2, and so on. We model the dynamics of the interactions as in \cref{eq:begin-measurement-procedure}: an imprint (a correlating operation) is followed by an inverse imprint (a readying/environmental correction operation). Thus, we have two kinds of interactions:
	\begin{enumerate}[label=(\roman*)]
		\item signal-environment interaction --- signal imprints on one of the environment qudits randomly followed by an inverse imprint by the qudit that succeeds it, and,
		\item environment-environment interaction --- one randomly chosen environment qudit interacts with its next neighbour, followed by an environment correction step by the neighbour that succeeds it.
	\end{enumerate}
	Further details are provided in \cref{app:sim_dets}.
	
	Motivated by the KL structure in \cref{sec:kl} and the microdynamics of \cref{sec:meas_proc}, we introduce a correlation metric \(C_\alpha\) intended to quantify how well each qudit \(\alpha\) ``agrees’’ with the remainder of the system in the pointer basis. This quantity is necessarily basis dependent and serves as a practical proxy for the degree to which \(\alpha\) participates in the formation of classical records. To define it, we begin with the \emph{pairwise agreement projector}
	\begin{equation} \label{eq:agreement_proj}
		\bbPi_{\alpha,\mu} := \sum_{k=0}^{d-1}{\ketbra{k}{k}_\alpha\otimes\ketbra{k}{k}_\mu},
	\end{equation}
	which projects onto the subspace where qudits \(\alpha\) and \(\mu\) share the same pointer value.  Next, we can define a ``signed match'' operator for the pair,
	\begin{equation}
		\mathbb{M}_{\alpha,\mu} := d\bbPi_{\alpha,\mu} - \mathbbm{1}_{\alpha,\mu}.
	\end{equation}
	where \(\mathbbm{1}_{\alpha,\mu} = \mathbbm{1}_{\alpha} \otimes \mathbbm{1}_{\mu}\) is the identity operator on the joint Hilbert space of qudits \(\alpha\) and \(\mu\). The operator \(\mathbb{M}_{\alpha,\mu}\) has eigenvalues \(d-1\) (when the qudits agree) and \(-1\) (when they disagree). Before turning to the full multi-qudit correlation measure, it is useful to illustrate the action of \(\mathbb{M}_{\alpha,\mu}\) on two-qudit states. For simplicity, we momentarily restrict to a state on \(\mathcal{H}_\alpha \otimes \mathcal{H}_\mu\). If the qudits are perfectly aligned in the pointer basis, \(\ket{\psi}_{\alpha\mu} = \sum_l \psi_l \ket{l}_\alpha \ket{l}_\mu\), then the agreement projector of \cref{eq:agreement_proj} acts as a stabilizer of the state, and in turn gives \(\mathbb{M}_{\alpha,\mu}\ket{\psi} = (d-1) \ket{\psi}\). On the other hand, for a generic separable two qudit state \(\ket{\psi\phi}_{\alpha\mu} = \sum_{k,l} \psi_k \phi_l \ket{k}_\alpha\ket{l}_\mu\), the expectation of the signed match operator is \(\braketOP{\psi\phi}{\mathbb{M}_{\alpha,\mu}}{\psi\phi}_{\alpha\mu} =  d \sum_k \abs{\psi_k}^2 \abs{\phi_k}^2 - 1\) which interpolates smoothly between \(d-1\) (perfect agreement) and \(-1\) (perfect misalignment). For an uncorrelated state, for example where all \(\psi_k\)s have around the same amplitude \(\approx \nicefrac{1}{\sqrt{d}}\), we find \(\braketOP{\psi\phi}{\mathbb{M}_{\alpha,\mu}}{\psi\phi}_{\alpha\mu} \approx 0\).
	
	We define the correlation metric for the qudit \(\alpha\) as the sum of contributions from pairs involving the system \(\alpha\),
	\begin{equation} \label{eq:Calpha_def}
		C_\alpha := \sum_{\mu \neq \alpha} \mathbb{M}_{\alpha,\mu},
	\end{equation}
	which aggregates the agreement of \(\alpha\) with every other subsystem in the pointer basis. By construction, for a global state \(\ket{\Psi} \in \mathcal{H}_{S} \otimes \mathcal{H}_{E}\), we have \(\braketOP{\Psi}{C_\alpha}{\Psi}_{SE}\) to be large when qudit \(\alpha\) belongs to a strongly correlated network of subsystems, and small when it is uncorrelated or misaligned. For an observer network state (such as that of \cref{eq:correlated-env-state-after-k-measurements}) in which a given qudit \(\alpha\) is aligned with \(K\) other qudits, the expectation value satisfies 
	\begin{equation}
		\braketOP*{\Psi}{C_\alpha}{\Psi}_{SE} \approx (d-1) K,
	\end{equation}
	reflecting the additive contribution of each agreeing pair. For the special case \(\alpha = 0\), we denote \(C_{S}:= C_0\), which quantifies how strongly the signal qudit correlates with the environment in the pointer basis.
	
	To quantify correlations within the environment as a whole, we define the
	\emph{environment correlation metric} as the sum over all environment-environment pairs,
	\begin{equation} \label{eq:CE_def}
		C_{{E}} := \frac{1}{2}\sum_{\alpha \in {E}} C_\alpha = \sum_{\substack{\alpha,\mu \in {E}\\ \alpha < \mu}} \mathbb{M}_{\alpha,\mu},
	\end{equation}
	where the factor of \(1/2\) removes double-counting of environmental qudits. The expectation value \(\braketOP{\Psi}{C_{{E}}}{\Psi}_{SE}\) therefore measures the total degree of internal agreement across the environment. For an environment containing a correlated block of size \(K\), one finds
	\begin{equation}
		\braketOP*{\Psi}{C_{{E}}}{\Psi}_{SE} \approx (d-1)\, \frac{K(K-1)}{2};
	\end{equation}
	it scales linearly with the number of agreeing pairs inside the correlated cluster.
	
	The interplay between \(C_{S}\) and \(C_{E}\) captures how the initial environmental correlation is converted into signal-environment correlation via the local imprint-type operations of \cref{sec:meas_proc}. When the environment state lies in a high-\(C_{E}\) subspace, many pairs of qudits are highly correlated. In this case, the environment is predisposed to be in a redundant low-entropy state. The system imprints its state via \(\mathcal{I}_{S \to E_1}\) and the environment corrects via \(\mathcal{I}^{-1}_{E_2 \to E_1}\), et cetera. Through this procedure, the observer network of the observer grows in size as measured by \(C_{S}\). On the other hand, if the initial state lies in a low-\(C_{E}\) subspace, the environment is uncorrelated or even anti-correlated: the environment does not have a strong pre-existing record, and there is no redundancy to begin with. Thus, it has a smaller capacity to transfer strong correlations to the system, and imprint operations do not affect the signal-environment correlation \(C_{S}\). For completeness, we also examine mutual information between the signal and environment fragments, which provides an independent diagnostic of redundant record formation.
	
	In our numerical simulations, we specialize to \emph{qubits} (\(d=2\)), so the global Hilbert space consists of one signal qubit and \(N\) environment qubits. The qubit versions of the correlation metrics and other details are discussed in \cref{app:correlation_ietric}. We run a numerical simulation in which a single signal qubit interacts unitarily with an \(N = 10\) qubit environment. We begin in an initially unentangled state of the signal and environment, \(\ket{\Psi^\text{init}}_{SE} = \ket{\psi}_{S}\otimes\ket{\chi}_{E}\). For concreteness, and since we primarily focus on the correlation capacity in the environment, we choose the signal to start in a uniform superposition \(\ket{\psi}_{S} = \big(\ket{0}_{S}  + \ket{1}_{S}\big)/\sqrt{2}\). The environment state \(\ket{\chi}_{E}\) is taken to be in states ranging from the maximally correlated state of \cref{eq:correlated-environment} to a fully generic Haar-random state, thereby sampling a wide range of initial environmental correlations. Interaction events, both signal-environment and environment-environment, occur stochastically in time, as detailed in \cref{app:sim_dets}. This framework allows us to monitor how correlation flows from the environment into the signal, and how the metrics \(C_{S}\) and \(C_{{E}}\) evolve under repeated local interactions.
	\begin{figure} [h!]
		\centering
		\includegraphics[width=0.8\textwidth]{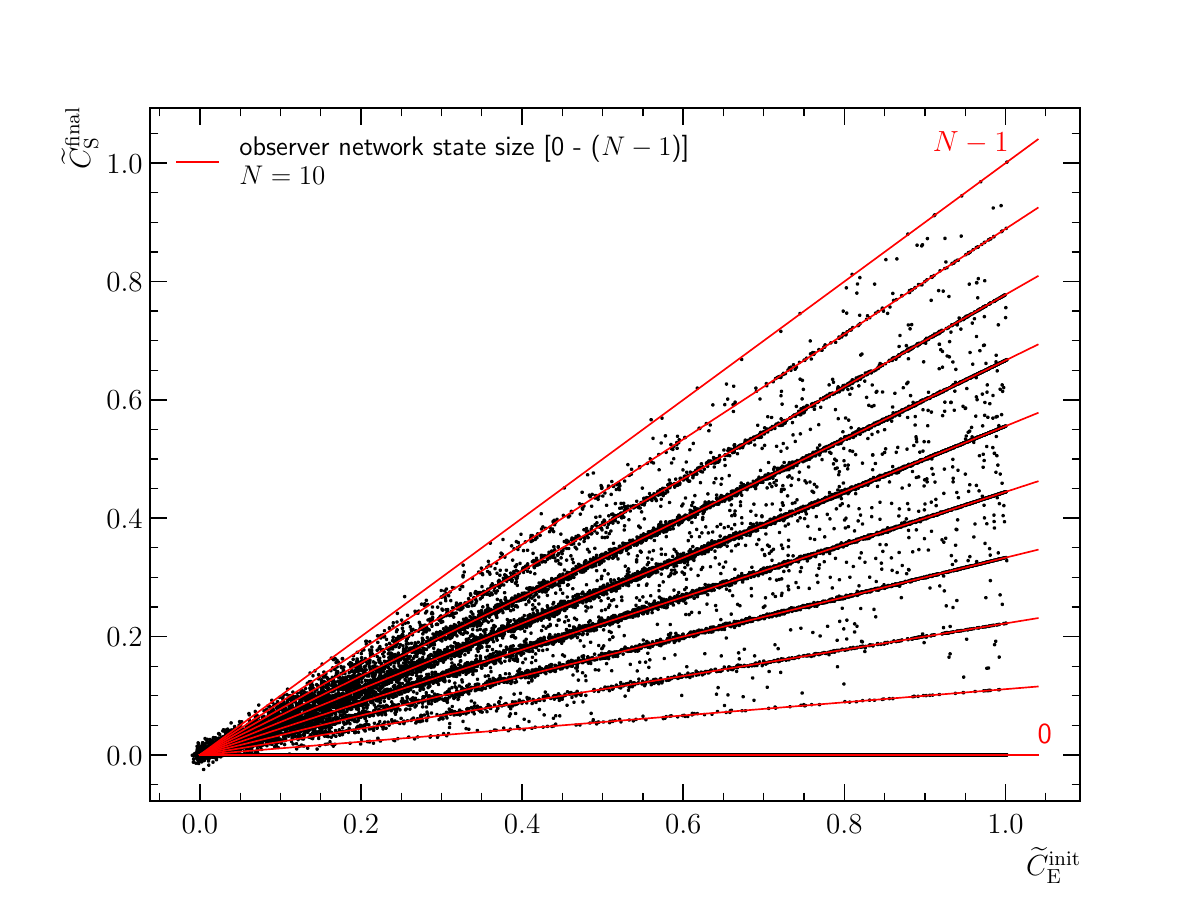}
		\caption{\label{fig:ce_cs} Larger initial correlation in the environment \(\widetilde{C}^\text{init}_{E}\) gives more final correlation \(\widetilde{C}^\text{final}_{S}\) in the resulting observer network state of the signal. The red lines above indicate the size of the emerging signal-observer correlation network. As detailed in \cref{sec:meas_proc}, the observer network for signal measurement extracts correlation from the environment ranging from 0 to \(N-1\): the initial information of the environment state must go somewhere. Thus, there needs to be at least one qubit to store this information. The values outside the lines are obtained because we take the average measure of \(\widetilde{C}^\text{final}_{S}\) over several time steps. If the measure changed over that time, an intermediate value is obtained. Details can be found in \cref{app:sim_dets}.}
	\end{figure}
	
	Since the signal starts in a uniform superposition and is initially unentangled with the environment, the signal correlation metric \(\braketOP{\Psi^\text{init}}{C_{S}}{\Psi^\text{init}}_{SE}\) vanishes. \(\braketOP{\Psi^\text{init}}{C_{E}}{\Psi^\text{init}}_{SE}\) captures the intrinsic correlation structure of the initial environment. For this reason, we compare the initial value \(\braketOP{\Psi^\text{init}}{C_{E}}{\Psi^\text{init}}_{SE}\) to the late-time value of \(\braketOP{\Psi^\text{final}}{C_{S}}{\Psi^\text{final}}_{SE}\), which measures the correlation the signal acquires once the imprint dynamics have saturated. \(\ket{\Psi^\text{final}}_{SE}\) is the final signal-environment state after the imprint dynamics have saturated. This acts as a measure of how much of the initial correlation is transferred to the signal-environment record. To compare correlation values across different runs and environment sizes, we work with the scaled versions of their expectation values,
	\begin{equation}
		\widetilde{C}_{S}^\text{final} := \frac{\braketOP*{\Psi^\text{final}}{C_{S}}{\Psi^\text{final}}_{SE}}{N-1}, \qquad 
		\widetilde{C}_E^\text{init} := \frac{\braketOP*{\Psi^\text{init}}{C_{E}}{\Psi^\text{init}}_{SE}}{\frac{1}{2}N(N-1)},
	\end{equation}
	which both lie in the interval \([0,1]\).
	
	As shown in \cref{fig:ce_cs} (see caption for more details), the final signal correlation \(\widetilde{C}_{S}^\text{final}\) grows monotonically with the initial environmental correlation \(\widetilde{C}_E^\text{init}\). Environments prepared close to fully correlated states (\(\widetilde{C}_E^\text{init} \approx 1\)) like those of \cref{eq:correlated-environment} produce large final observer networks in
	which about half of the environmental fragments agree with the signal, yielding \(\widetilde{C}_{S}^\text{final}\) averaging around 1/2. In contrast, for Haar-random environment states, the expectation value of each pairwise match operator \(\mathbb{M}_{\alpha\mu}\) vanishes, so the scaled environmental correlation \(\widetilde{C}_E^\text{init}\) concentrates sharply near zero. Thus, random states represent environments with essentially no pointer-basis agreement; therefore, the signal never acquires a stable observer network, and the final \(\widetilde{C}_{S}^\text{final}\) remains close to zero.
	\begin{figure} [h!]
		\centering
		\includegraphics[width=0.8\textwidth]{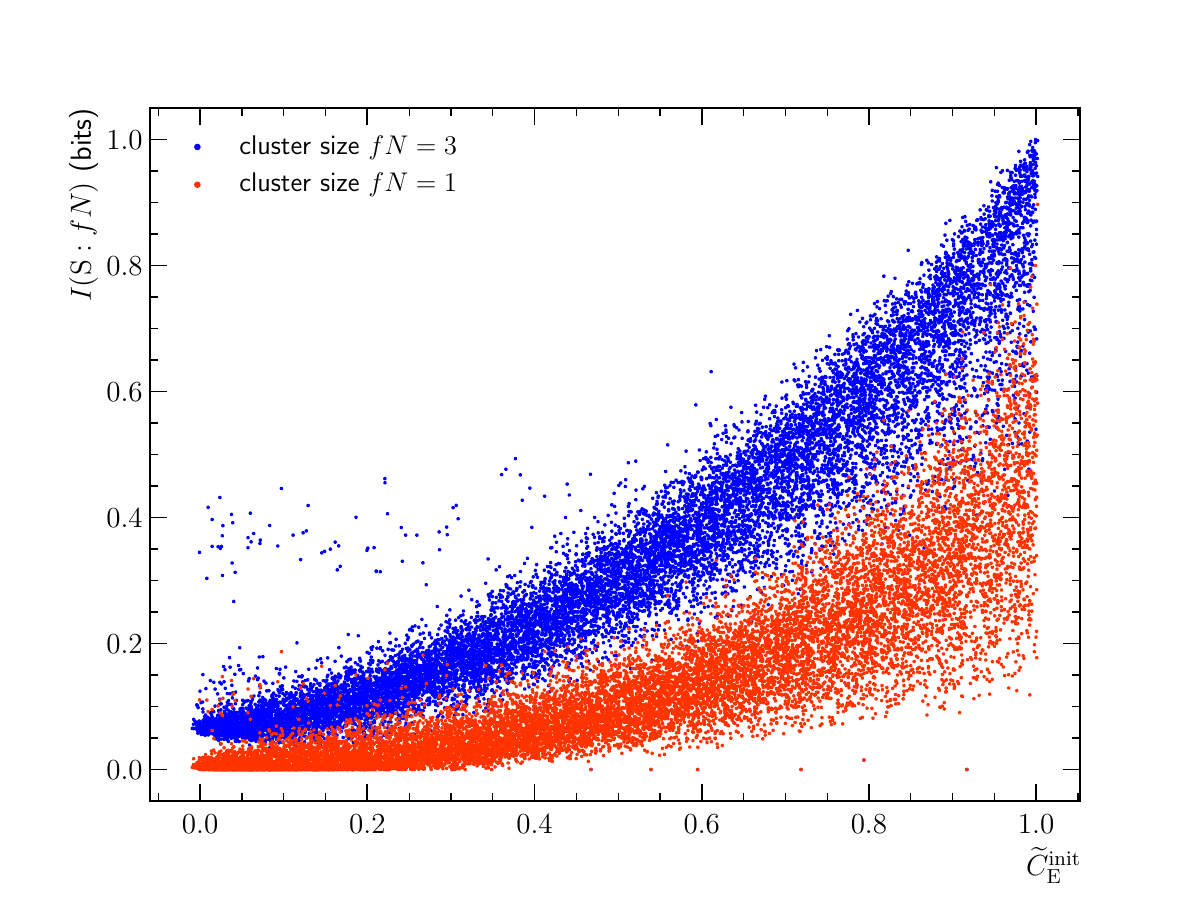}
		\caption{\label{fig:ce_iut} Initial environment states with larger initial correlation \(\widetilde{C}_E^\text{init}\) give rise to final observer network states with larger mutual information between signal and environment. Larger cluster sizes (red) yield more mutual information on average, as expected.}
	\end{figure}
	
	As an independent check, we compute the mutual information \(I({S}:fN)\) between the signal and clusters of size \(fN\) (for a fraction \(f\) of the \(N\) qubits) of the environment, a standard diagnostic in the quantum Darwinism literature. Mutual information plays a central role in quantifying how information about a system becomes distributed across its environment by measuring the amount of knowledge an observer can acquire by accessing fragments of the environment. As shown in \cref{fig:ce_iut}, the mutual information exhibits the same overall trend: environments with larger initial \(\widetilde{C}_E^\text{init}\) give rise to environmental fragments with larger information about the signal. Moreover, larger clusters capture more information on average than smaller clusters, as expected. While the precise shape of the scatter depends on the details of our sampling of initial states, the rising trend is robust: initial environmental correlation enhances the information captured by environmental fragments.
	\begin{figure} [h!]
		\centering
		\includegraphics[width=0.8\textwidth]{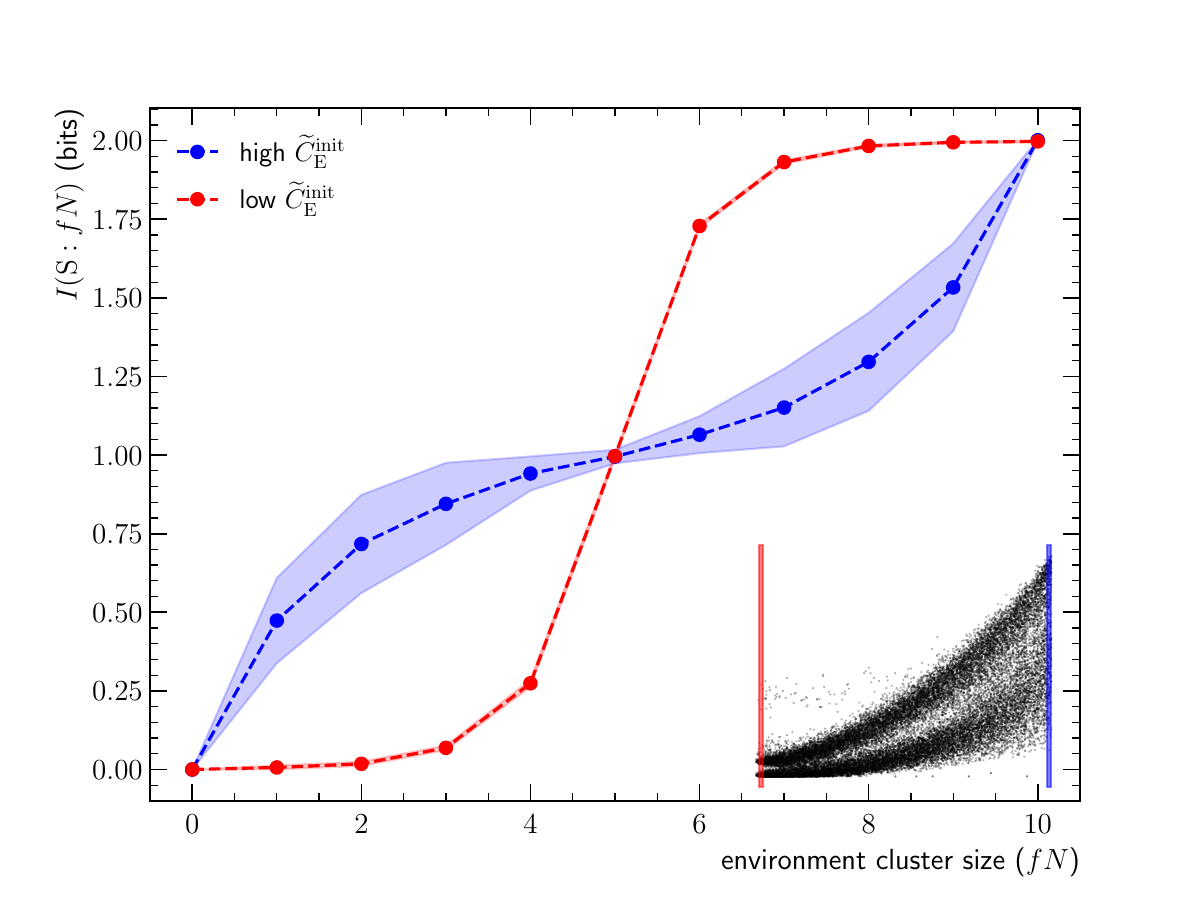}
		\caption{\label{fig:mut_cluster} Mutual information (and its one-sigma deviations) plotted with respect to cluster size shows interesting but expected trends. For states with low initial correlation (red), the mutual information does not show a classicality plateau. On the other hand, states with high initial correlation (blue) exhibit a classicality plateau. Inset in the lower right corner is \cref{fig:ce_iut}. The data points contributing to the blue and red curves above are sampled from the points lying within the blue and red regions in the inset figure, respectively.}
	\end{figure}
	
	Finally, \cref{fig:mut_cluster} shows the mutual information \(I({S}:fN)\) as a function of cluster size for two classes of initial environment states --- those with large initial correlation \(\widetilde{C}_E^\text{init}\) (blue) and those with small \(\widetilde{C}_E^\text{init}\) (red). As emphasized in article \cite{riedel-zurek}, the behavior of \(I({S}:fN)\) with fragment size provides a direct probe of how information about the system spreads through the environment. For highly correlated initial environments, we observe the characteristic \emph{classicality redundancy plateau} \cite{riedel-zurek, Blume_Kohout_2006}, where even small clusters contain nearly the complete classical information about the signal, indicating that many disjoint fragments independently carry the same record. By contrast, environments with low initial correlation show no such plateau. The mutual information increases only gradually with fragment size, reflecting that information about the signal is stored in a non-redundant, globally distributed way. Thus, such environments lack the structured correlations required to broadcast the signal's pointer value into multiple fragments.
	
	These findings confirm the theoretical picture developed in \crefrange{sec:corr_res}{sec:meas_proc} that environmental correlation functions as the \emph{operational resource} that enables the unitary measurement dynamics to produce robust, redundant records for emergent classicality.
	
	\section{Discussion} \label{sec:discussion}
	\subsubsection{Summary:}
	In this work, we have shown that correlation in the environment functions as a finite physical resource that enables quantum measurement. Beginning from a set of physically natural assumptions: (i) universal unitary evolution, (ii) a unitary measurement interaction that produces distinct pointer outcomes (\cref{eq:quantum_ieasurement}), (iii) orthogonality of the corresponding environmental branches, and (iv) finite Hilbert-space dimensions, we proved a no-go theorem demonstrating that these statements are incompatible if the environment is allowed to begin in an arbitrary state. The theorem implies that successful, objective measurement is possible only when the environment initially lies within a restricted subspace \(\Phi\), whose structure and dimensionality limit its capacity to form redundant classical records. Interpreting this constraint operationally leads to a resource inequality, \(K\ln(d) + \ln(\dim\Phi) \le \ln(\dim \mathcal{H}_E)\), which quantifies how the ability to produce \(K\) redundant records of a \(d\)-outcome measurement is determined by the size of the initial subspace \(\Phi\). We further showed that enforcing perfect distinguishability of conditional environmental states forces the measurement maps to satisfy the Knill-Laflamme condition of quantum error correction, thereby inducing a natural tensor-factorisation of the relevant post-measurement subspace into a ``pointer'' factor, which stores the outcome, and a ``memory'' (of the prior environment) factor, which retains the environment’s prior information in accordance with the no-deletion theorem. This also opens a new line of inquiry connecting quantum error correction with the robust, redundant records formed à la quantum Darwinism, and we hope to build on this connection in future work.
	
	To show that the algebraic structure derived in \cref{sec:corr_res,sec:kl} can arise from ordinary quantum dynamics, we constructed in \cref{sec:meas_proc} an explicit qudit model based on spacetime-local imprint and inverse-imprint interactions. With the environment prepared in the fully correlated state of \cref{eq:correlated-environment}, these local unitaries implement the two steps identified abstractly in \cref{sec:kl}: a readying step that compresses the prior environment information into a memory sector (in accordance with no-deletion) while aligning several fragments into a common pointer ``ready'' state, and a correlating step that redundantly imprints the signal pointer value across multiple environmental qudits, producing an explicit observer-network state. Numerical simulations of this model  (\cref{sec:num_res}) corroborate the theoretical picture: environments with greater initial correlation yield higher-fidelity, more redundant records. Together, these results demonstrate that correlation is a finite, consumable resource whose transfer from environment to system underlies the emergence of objective classical reality within unitary quantum mechanics. 
	
	We now turn to several natural extensions and open questions that arise from our analysis. These include relaxing different statements part of the no-go theorem of \cref{sec:corr_res}, the robustness of record formation under basis misalignment, and possible dynamical origins of the correlated environmental reserves required for measurement.
	
	\subsubsection{Relaxing conditions of the no-go theorem:}
	We take a specific stance in this work: we abandon arbitrary initial states of the environment and demonstrate how to make quantum measurement work in that context. What happens when we give up on the other \crefrange{stat:universality}{stat:finiteness}? Let us begin with \cref{stat:finiteness}: in the case of a (countably) infinite-dimensional Hilbert space, one can use the Hilbert hotel argument to accommodate the infinite numbers in the initial state of the environment, along with some additional information. We show this using a simple counterexample. Consider the initial state of the environment as \(\ket{\phi}_{E} = \sum_{i=0}^{\infty}{\phi_i\ket{i}_{E}}\) and a signal system as \(\ket{\psi}_{S} = \psi_0\ket{0}_{S} + \psi_1\ket{1}_{S}\). In this case, we can define a measurement procedure as
	\begin{equation}
		\mathcal{U}^\text{meas} \ket{\psi}_{S}\ket{\phi}_{E} = \sum_{i=0}^{\infty}{\phi_i \left(\psi_0\ket{0}_{S}\ket{2 \times i}_{E} + \psi_1\ket{1}_{S}\ket{2 \times i+1}_{E}\right)}.
	\end{equation}
	Though contrived, it would satisfy \cref{stat:universality,stat:measurement,stat:orthogonality,stat:arbitrary}. Thus, dealing with infinite-dimensional Hilbert spaces requires more careful consideration.
	
	Next, we consider giving up on \cref{stat:orthogonality} --- in this regard consider orthogonality only upto \(\epsilon\): \(\braket{\mathcal{E}_{i\phi}}{\mathcal{E}_{j\phi}} = (1-\epsilon)\delta_{ij} + \mathcal{O}(\epsilon)\). We expect to get weaker bounds on the dimensionality constraint of \cref{eq:dimension_inequality}: one can fit in more records for the same balance between \(\dim\mathcal{H}_{E}\) and \(\dim\Phi\). In general, we expect a trade-off between the number of records that can be generated and the level of orthogonality of environmental records. We do not establish a precise connection in this work, but refer to other works \cite{Brand_o_2015,PhysRevLett.118.150501,PhysRevLett.122.010403} as a starting point for future research.
	
	If we abandon \cref{stat:universality,stat:measurement}, there would be no requirement for information conservation, as wavefunction collapse involves a loss of information (for example, see \cite{RevModPhys.85.471, PhysRevD.34.470}). As a result, there would be no constraints on the initial state of the environment. Once again, we do not make a precise connection in this work.
	
	\subsubsection{Emergence of correlation:}
\begin{figure} [h!]
		\centering
		\includegraphics[width=0.8\textwidth]{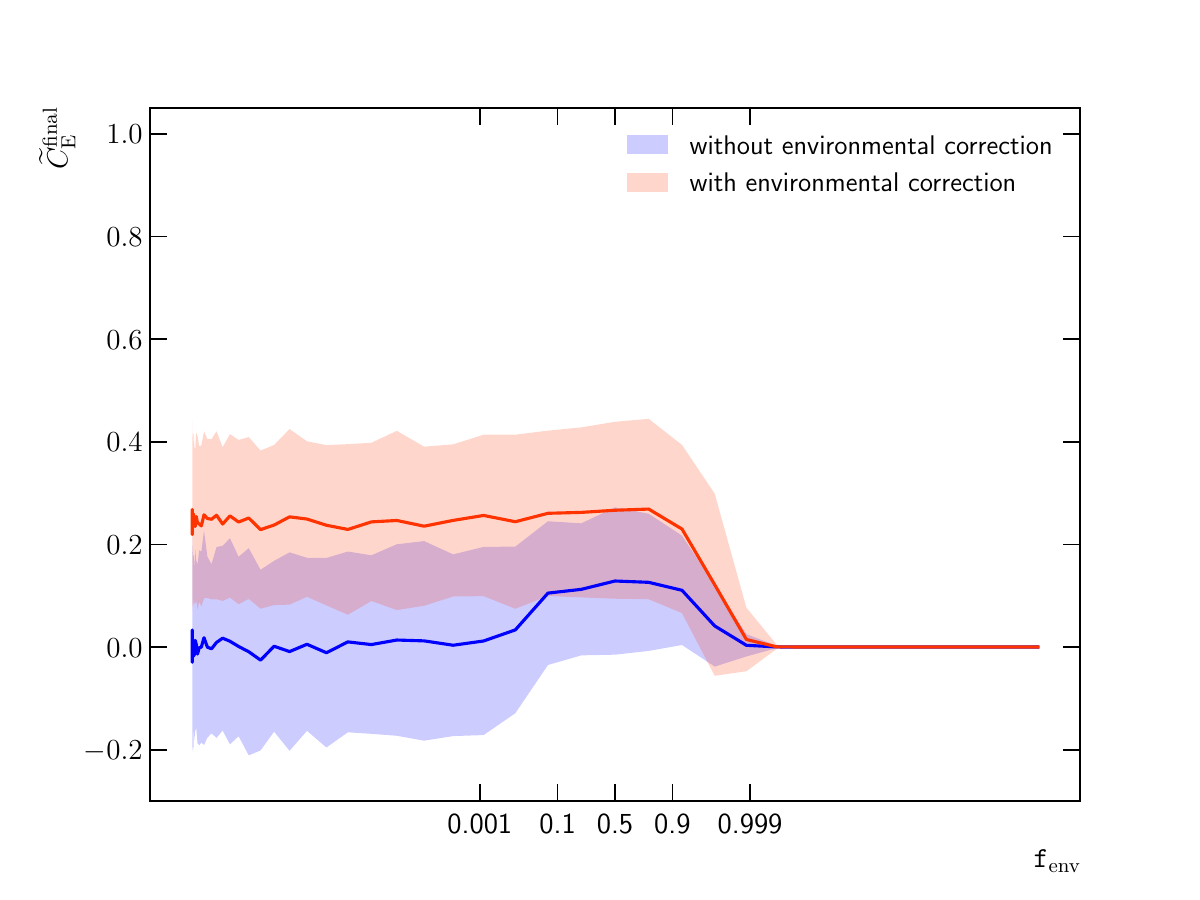}
		\caption{\label{fig:fracenv} Variation of the scaled correlation metric for signal (\(\widetilde{C}^\text{final}_{S}\)) as a function of the fraction of total interactions that are between environmental qubits (\(\mathtt{f}_{env}\), see \cref{app:sim_dets}). The red curve represents quantum dynamics with environmental correction, as described in \cref{sec:meas_proc}. The blue curve represents quantum dynamics without environmental correction, see \cref{app:sim_dets}. We show the average over many runs along with the one standard deviation boundaries. Even without environmental correction, a particular resonance between environmental-environmental interactions and signal-environmental interactions (\(\mathtt{f}_{env}\)) produces reliable records.}
	\end{figure}

	If correlation is a finite resource required for measurement, a natural question is how such correlated environmental reserves arise in the first place. While a complete explanation is beyond the scope of this work, our numerical simulations suggest a plausible dynamical mechanism. As shown in \cref{fig:fracenv}, record formation in the local imprint model depends sensitively on the balance between signal-environment interactions, which generate correlations, and environment-environment interactions, which tend to redistribute or erase them. Even with explicit environmental correction switched off in the simulation, which would otherwise perform the ``readying'' step, we observe a resonance regime in which an intermediate fraction of environment-environment interactions spontaneously produces states capable of supporting reliable records (blue curve). Introducing environmental correction again amplifies and stabilizes this regime (red curve), significantly increasing the environment's capacity to supply correlation.
	
	These observations motivate the hypothesis that correlated environments can arise generically in complex quantum systems through a form of dynamical selection: interaction patterns that accidentally support stable record formation accumulate and persist, while those that rapidly wash out correlations do not. We hypothesise that such mechanisms were involved in the formation of ``classical''  systems whose state is redundantly recorded; for example, biological systems. We hypothesise that these biological systems were initially tuned to this resonance and utilized it to enable the formation of correlated records --- internal representations within them of external signals. This allowed them to react to the environment and become fitter. Later, these mechanisms became more robust through biological evolution via ``environmental correction'', allowing for the extraction of correlations more effectively, as we propose in \cref{sec:meas_proc}. In general, we hypothesise that there were two eras in the emergence of systems that support the creation of \emph{classical records}: the past, dominated by the dynamics of the blue curve, and the present, dominated by the dynamics in the red curve after evolving mechanisms of environmental correction (\cref{fig:fracenv}). Although speculative, this dynamical viewpoint offers a potential route toward understanding how the correlation required for measurement might itself emerge from the underlying quantum dynamics of the universe (or parts of it).
	
	\subsubsection{Measurement in different bases:}
	Our analysis so far has assumed that all environmental fragments measure the signal in the same pointer basis. It is therefore natural to ask how robust the resulting objectivity is to small misalignments of this basis. Suppose the signal system is expanded in a basis \(\{\ket{i}\}\). One set of environmental fragments \(E\) is correlated in this basis, and another set \(F\) is correlated in a slightly rotated basis \(\{\ket{i'}\}\). A unitary transformation relates the two bases,
	\begin{equation}
		\ket{i'}=\sum_i \mathsf{U}^\dagger_{i'i}\ket{i}, \qquad 
		\ket{i}=\sum_{i'} \mathsf{U}_{ii'}\ket{i'}.
	\end{equation}
	First, the signal interacts with \(E\) and forms an observer-network state; it then interacts with \(F\), producing a second network. The combined evolution is
	\begin{subequations}
		\begin{alignat}{2}
			\sum_i \psi_i \ket{i}_{S}
			\overset{E\text{ measures}}{\longrightarrow}& 
			& &\sum_i \psi_i \ket{i}_{S}\ket{ii\cdots}_{E} \\
			& & =&\sum_{i j'} \psi_i\, \mathsf{U}_{i j'} \ket{j'}_{S}\ket{ii\cdots}_{E} \\
			\overset{F\text{ measures}}{\longrightarrow}&
			& &\sum_{i j'} \psi_i\, \mathsf{U}_{i j'} \ket{j'}_{S}\ket{ii\cdots}_{E}\ket{j'j'\cdots}_{F}.
			\label{eq:non_aligned_ieasuements}
		\end{alignat}
	\end{subequations}
	\begin{figure} [h!]
		\centering
		\begin{subfigure}{0.34\textwidth}
			\centering
			\begin{tikzpicture}
				\draw[->] (-0.5,0) -- (2.5,0) node[below right] {\(2\)};
				\draw[->] (0,-0.5) -- (0,2.5) node[above] {\(0\)};
				\draw[->] (0.25,0.25) -- (-1,-1) node [below left] {\(1\)};
				
				\draw[->, rotate around={10:(0,0)}] (-0.5,0) -- (2.5,0) node[below right] {\(2'\)};
				\draw[->, rotate around={10:(0,0)}] (0,-0.5) -- (0,2.5) node[above] {\(0'\)};
				\draw[->, rotate around={10:(0,0)}] (0.25,0.25) -- (-1,-1) node [below left] {\(1'\)};
			\end{tikzpicture}
			\caption{\label{fig:bases_rotated} Rotated bases.}
		\end{subfigure}
		\hfill
		\begin{subfigure}{0.64\textwidth}
			\centering
			\begin{tikzpicture}[domain=-1.5:1.5]
				\draw[<->] (-3.2,0) -- (3.2,0) node[right] {\(\epsilon\)};
				\draw[->] (0,-0.3) -- (0,2.25) node[right] {\(\mathcal{E}(\epsilon)\)};
				\draw[-] (-0.1,2) -- (0.1,2) node[right] {1};
				
				\draw[smooth] plot (2*\x, {2*2*sin(\x r)*sin(\x r)*cos(\x r)*cos(\x r)});
				\draw[smooth, dashed, blue] plot (\x, {\x*\x});
				
				\draw[smooth] (-3.2,-1) -- (-2.8,-1) node[right] {- Actual probability distribution};
				\draw[smooth, dashed, blue, local bounding box = A] (-3.2,-1.5) -- (-2.8,-1.5);
				\node[right of = A,xshift=1cm] {- Square approximation};
				\draw[thick] (-3.1,0.15) -- (-3.1,-0.15) node[below] {\(-\pi\)};
				\draw[thick] (3.1,0.15) -- (3.1,-0.15) node[below] {\(+\pi\)};
			\end{tikzpicture}
			\caption{\label{fig:basis-mismatch-scaling} Error probability.}
		\end{subfigure}
		\caption{\textbf{(a)} Two bases \(\{i\}\) and \(\{i'\}\) are schematically represented above. Note that we schematically only represent a subset \(\{0,1,2\}\subseteq \{i\}\) of the label set. Each label \(i\) matches the label \(i'\), and they are shifted with respect to each other by a small angle \(\epsilon\). In case the angle is small outcomes \(i\) and \(i'\) match each other with a very high probability: \(1-\mathcal{O}(\epsilon^2)\). Mismatched outcomes are observed with a probability \(\mathcal{O}(\epsilon^2)\). \textbf{(b)} (\textit{Error probability of misaligned measurements}) Observer \(E\) measures the signal in the basis \(\{\ket{0}, \ket{1}\}\). Following this, observer \(F\) measures the signal in the rotated basis \(\{\ket{0'}, \ket{1'}\}\) specified by the unitary \(\mathsf{U}(\epsilon) =( \cos(\epsilon)\ket{0'} + \sin(\epsilon)\ket{1'})\bra{0} + (-\sin(\epsilon)\ket{0'} + \cos(\epsilon)\ket{1'})\bra{1}\). The probability of error, \(\mathcal{E}(\epsilon)\), for the observers \(E\) and \(F\) to disagree with one another, is the probability for the branches \(\ket{00\dots}_{E}\ket{1'1'\dots}_{SF}\) or \(\ket{11\dots}_{E}\ket{0'0'\dots}_{SF}\) to be realised: \(\mathcal{E}(\epsilon) = \sin(\epsilon)^2\). We note that for small \(\epsilon\), the probability of error scales as \(\epsilon^2\).}
	\end{figure}
	To study small mis-alignments, we parameterize the rotation as \(\mathsf{U} = e^{\epsilon \mathsf{A}} = \mathbbm{1} + \epsilon \mathsf{A} + O(\epsilon^2)\), where \(\epsilon\ll 1\) and \(\mathsf{A}\) is anti-Hermitian. Identifying the labels \(i\) and \(i'\) when the bases differ only by the small angle \(\epsilon\) (see \cref{fig:bases_rotated}), we expand \cref{eq:non_aligned_ieasuements} to obtain
	\begin{equation}
		\sum_i \psi_i \ket{i'}_{S}\ket{ii\cdots}_{E}\ket{i'i'\cdots}_{F}
		+
		\epsilon \sum_{i j'} \psi_i \mathsf{A}_{i j'} 
		\ket{j'}_{S}\ket{ii\cdots}_{E}\ket{j'j'\cdots}_{F}
		+ O(\epsilon^2).
	\end{equation}
	
	The leading term describes perfect agreement between the two observer networks: both record the same outcome \(i\). The next-order term produces branches in which the two networks disagree, but with probability proportional to \(\epsilon^2\). This scaling, illustrated explicitly in \cref{fig:basis-mismatch-scaling}, shows that small basis misalignments produce only second-order errors. Thus, objective classical reality, with redundant and robust records across multiple environmental fragments, does not require exact alignment of the measurement bases. Instead, it remains stable under small deviations, with the mismatch probability being proportional to the square of the misalignment angle. This is consistent with the picture developed earlier, in which the preferred basis is determined by the structure of the environment and its correlations; small perturbations away from this basis do not disrupt the redundancy of the records formed.
	
	Our discussion here does not fully resolve the broader preferred-basis problem \cite{PhysRevD.26.1862, PhysRevD.24.1516}, nor does it address how the relevant pointer basis emerges dynamically. However, these results indicate that once a correlated environment selects a pointer basis, high-fidelity record formation remains robust under small variations. This robustness also resonates with ideas in \emph{quantum mereology} \cite{Carroll_2021, Loizeau2023, Loizeau2024, Adil2024, Zanardi_2004}, where the very identification of subsystems, and hence of the emergent classical degrees of freedom in which records are stored, arises from the structure of correlations and its dynamics rather than being fixed a priori.
	
	Extending our numerical analysis to higher-dimensional qudits and developing quantitative measures of basis stability in that setting are interesting directions for future work. Additionally, the possibility of probing such effects experimentally or in cosmological contexts is also worth exploring. Another direction would be to consider how such an approach can be used to understand the emergence of biological perception systems in early life.
	
	\section*{References}
	\bibliographystyle{utphys}
	\bibliography{references}
	
	\begin{appendices}
		\crefalias{section}{appendix}
		\crefalias{subsection}{appendix}
		\crefalias{subsubsection}{appendix}
		\section{The Correlation Metric} \label{app:correlation_ietric}
		In this appendix, we give further context about the correlation metric \(C_\alpha\) and \(C_{E}\) introduced in \cref{sec:num_res} and fill in some technical details. We start with a correlation metric for qubits. 
		
		\subsubsection{Correlation metric for qubits:}
		For \(d=2\) (qubits), one has \(\bbPi_{\alpha,\mu} = \ketbra{0}{0}_\alpha\otimes\ketbra{0}{0}_\mu + \ketbra{1}{1}_\alpha\otimes\ketbra{1}{1}_\mu\) and so
		\begin{equation}
			2\bbPi_{\alpha,\mu} - \mathbbm{1} = Z_\alpha Z_\mu,
		\end{equation}
		where \(Z\) is the Pauli \(Z\) operator. The system-relative correlation is
		\begin{equation}
			C_\alpha = Z_\alpha \sum_{\mu\neq \alpha} Z_\mu.
		\end{equation}
		
		The states for \(N\) qubits (\(E_1\) - \(E_N\)) can be enumerated with  index \(I = 0,\dots,2^N-1\). Then the state \(\ket{I}\) (with an implicit \(N\)) can be defined to correspond to the state where the environment \(E_\alpha\) has the value of the \(\alpha\)th term in the binary expansion of \(I\). Thus, we get (for \(N=3\), for example) \[\ket{0}_{E} = \ket{0}_{E_1}\ket{0}_{E_2}\ket{0}_{E_3}, \qquad \ket{1}_{E} = \ket{0}_{E_1}\ket{0}_{E_2}\ket{1}_{E_3},\dots,\qquad \ket{5}_{E} = \ket{1}_{E_1}\ket{0}_{E_2}\ket{1}_{E_2}, \dots.\]
		
		When the signal starts at the state \(\ket{\psi}_{S} = \nicefrac{1}{\sqrt{2}}\ket{0}_{S} + \nicefrac{1}{\sqrt{2}}\ket{1}_{S}\) and goes through the measurement procedure, the general overall state of the signal-environment system takes the form:
		\begin{equation}
			\ket{\Psi}_{SE} = \frac{1}{\sqrt{2}}\ket{0}_{S}\ket{E_0}_{E} + \frac{1}{\sqrt{2}}\ket{1}_{S}\ket{E_1}_{E},
		\end{equation}
		where the conditional environment states are \(\ket{E_0} = \sum_I c_I\ket{I}_{E}\) and \(\ket{E_1} = \sum_I \bar{c}_I\ket{I}_{E}\) for coefficients \(c_I\) and \(\bar{c}_I\) respectively. It is seen that
		\begin{equation}
			C_{S} \ket{0}_{S}\ket{I}_{E} = \text{Hamm}(I) \ket{0}_{S}\ket{I}_{E} \quad \text{and} \quad C_{S} \ket{1}_{S}\ket{I}_{E} = -\text{Hamm}(I) \ket{1}_{S}\ket{I}_{E}
		\end{equation}
		where \(\text{Hamm}(I)\) is the \text{Hamm}ing weight of the binary expression for \(I\) --- the number of 0s minus the number of 1s. Thus, \(\braketOP{\Psi}{C_{S}}{\Psi}_{SE}\) evaluated would give
		\begin{equation}
			\frac{1}{2} \sum_I{(\abs{c_I}^2 - \abs{\bar{c}_I}^2)\text{Hamm}(I)}.
		\end{equation}
		
		\subsubsection{Correlation metric in other circumstances:}
		As a concrete, more elaborate example, we consider four qutrits (\(N = 4, d = 3\)) labelled \(\mu = 0,1,2,3\). We get for the environmental correlation
		\begin{equation}
			C_{E} = d \sum_{\mu<\nu} \bbPi_{\mu,\nu} - \binom{4}{2} \mathbbm{1}	= 3 \sum_{\mu<\nu} \bbPi_{\mu,\nu} - 6 \mathbbm{1},
		\end{equation}
		where 
		\begin{equation}
			\bbPi_{\mu,\nu} = \sum_{i=1}^{3} \ketbra{m}{m}_\mu \otimes \ketbra{m}{m}_\nu.
		\end{equation}
		Below, we provide three example states: a GHZ-type state (maximally correlated), a uniform product state (zero correlation), and a ``cyclic-anti'' state (negative correlation).
		
		\emph{\uline{Positively correlated state:}} a GHZ-like state
		\begin{equation}
			\ket{\Psi_\text{GHZ}} = \frac{1}{\sqrt{3}}\left(\ket{0000}+\ket{1111}+\ket{2222}\right).
		\end{equation}
		Any pair of qudits \(\mu,\nu\) matches with probability 1. There are \(\binom{4}{2}=6\) unordered pairs, so the total match is \(6\). Multiplying by \(d=3\) yields \(18\), and then we subtract \(6\). Hence,
		\begin{equation}
			C_{E}(\ket{\Psi_\text{GHZ}}) = 3\times6 - 6 = 12.
		\end{equation}
		All pairs are perfectly aligned, yielding a large positive value, which in this case is at its maximum.
		
		\emph{\uline{Uncorrelated state:}} each qudit is in a uniform superposition
		\begin{equation}
			\ket{\Psi_0} = \left(\frac{1}{\sqrt{3}}(\ket{0}+\ket{1}+\ket{2})\right)^{\otimes4}.
		\end{equation}
		Each pair of qudits has a \(\nicefrac{1}{3}\) probability of matching. Summing over 6 pairs gives \(6\times\nicefrac{1}{3}=2\). Multiply by 3 to get 6, then subtract 6, so
		\begin{equation}
			C_{E}(\ket{\Psi_0}) = 6-6 = 0.
		\end{equation}
		The state is uncorrelated beyond random chance, so the operator yields zero.
		
		\emph{\uline{Anti-correlated state:}} consider
		\begin{equation}
			\ket{\Psi_\text{anti}} = \frac{1}{\sqrt{3}} \left(\ket{0120} + \ket{1201} + \ket{2102}\right).
		\end{equation}
		In each branch, exactly one pair of qudits (namely \(\mu = 0\) and \(\nu = 3\)) matches, while the other five pairs do not match. Hence, the total match is 1. Multiplying by 3 yields 3, then we subtract 6, giving
		\begin{equation}
			C_{E}(\ket{\Psi_\text{anti}}) = 3-6 = -3.
		\end{equation}
		Most pairs are misaligned, resulting in a total correlation below the random baseline and a negative value.
		
		\section{Details of the Simulation} \label{app:sim_dets}
		In this appendix, we give details about the simulation procedure used. The big picture is that a signal system \(S\) interacts and transfers information about its state with environment systems in a bath, \(E_1\) - \(E_N\). The environment systems themselves interact with one another and may serve to erase the signal record or to propagate it, depending on the parameters used.
		
		Signal and environment systems are qubits. Our aim with the simulation is to show that large amounts of correlation in the environment (\(C_{E}\)) lead to large amounts of correlation in the signal observer network (\(C_{S}\)). We list the simulation parameters individually.
		
		\subsubsection{Initial state:}
		The signal system always begins in the state:
		\begin{equation}
			\ket{\psi}_{S} = \frac{1}{\sqrt{2}} \ket{0}_{S}+ \frac{1}{\sqrt{2}} \ket{1}_{S}. 
		\end{equation}
		
		To explore the relevant part of the Hilbert space for our hypothesis, we use two ingredients:
		\begin{enumerate}
			\item a correlated state --- a GHZ like state of the environment with maximal \(C_{E}\), \(\ket{\text{corr}}_{E} = \alpha\ket{00\dots0}_{E} + \beta\alpha\ket{11\dots1}_{E}\), where \(\alpha\) and \(\beta\) are complex numbers with \(\abs{\alpha}^2 + \abs{\beta}^2 = 1\), and,
			\item a Haar random state --- a Haar random state from the entire Hilbert space \(\mathcal{H}_{E}\), \(\ket{\text{Haar}}_{E}\).
		\end{enumerate}
		The states we use to explore our hypothesis are superpositions of the corr and Haar states:
		\begin{equation}
			\ket{\phi}_{E} = \alpha'\ket{\text{corr}}_{E} + \beta'\ket{\text{Haar}}_{E}
		\end{equation}
		where \(\abs{\alpha'}^2 + \abs{\beta'}^2 = 1\).
		
		While the states explored do not sample the Hilbert space uniformly, as a Haar measure does, they efficiently sample the states we require to test our hypothesis. Next, we examine the interactions used.
		
		\subsubsection{Interactions:}
		The interactions in the case of qubits are simple:
		\begin{itemize}
			\item \(\mathcal{I}_{\alpha \to \mu}\): \(\texttt{CNOT}_{\alpha, \mu}\), a controlled not gate between qubits \(\alpha\) and \(\mu\),
			\item \(\mathcal{I}^{-1}_{\alpha \to \mu}\): \(\texttt{CNOT}_{\alpha, \mu}\), also a controlled not gate between qubits \(\alpha\) and \(\mu\) as the \texttt{CNOT} is its own inverse.
		\end{itemize}
		
		The signal \(S\) interacts randomly with various environment qubits. On the other hand, we restrict the environment qubits to only interact with their nearest neighbours --- the qubits \(E_1\) to \(E_N\) form a chain with the last qubit the neighbour of the first. In all the following, addition is modulo N so that the environments form a ring. Signal is indicated by index 0. Thus, the qudit measurement procedure uses the interactions:
		\begin{itemize}
			\item \(\texttt{CNOT}_{\alpha+1,\alpha}\circ\texttt{CNOT}_{0,\alpha}\) for signal-environment interactions, and,
			\item \(\texttt{CNOT}_{\alpha+1,\alpha}\circ\texttt{CNOT}_{\alpha-1,\alpha}\) for environment-environment interactions.
		\end{itemize}
		
		\emph{Alternative interaction:} we found that a certain resonance between the signal-environment and environment-environment interactions allows the spontaneous emergence of classical records (see \cref{fig:fracenv}). For this, we use the alternative interactions without the correction terms:
		\begin{itemize}
			\item \(\texttt{CNOT}_{0,\alpha}\) for signal-environment interactions, and,
			\item \(\texttt{CNOT}_{\alpha-1,\alpha}\) for environment-environment interactions.
		\end{itemize}
		
		\subsubsection{Time behaviour:}
		As the interactions are random events, we need to model the times of events explicitly. We assume that the events are Poisson distributed over a continuous time axis. This would mean that the waiting times between interactions can be efficiently modelled using an exponential distribution, which we do in our code. The various parameters used are:
		\begin{itemize}
			\item \(\mathtt{T}_\text{int}\) ---  the average time between interactions (a parameter of the exponential distribution) set to 1,
			\item \(\mathtt{N}_\text{int}\) --- the number of interactions, set to 200,
			\item \(\mathtt{T}_\text{flyby}\) --- the time for which the signal interacts with the environment qubits, set to \(0.3\times\mathtt{T}_\text{int}\times\mathtt{N}_\text{int} = 60\) (see \cref{fig:time_behaviour}).
		\end{itemize}
		
		At each interaction, a random number determines whether the interaction will be a signal-environment interaction or an environment-environment interaction. A parameter \(\mathtt{f}_\text{env}\) sets the ratio:
		\begin{equation}
			\mathtt{f}_\text{env} = \frac{\text{\# environment-environment interactions}}{\text{\# total interactions}}.
		\end{equation}
		The following values are used for \(\mathtt{f}_{env}\):\\
		\begin{tabularx}{0.3\textwidth}{XX}
			\cref{fig:ce_cs} & \(\in [0,1]\), \\
			\cref{fig:ce_iut} & \(\lesssim 0.023\), \\
			\cref{fig:mut_cluster} & \(\sim 0\), \\
			\cref{fig:time_behaviour} & \(8.2e-3\).
		\end{tabularx} \\
		The value is varied for \cref{fig:fracenv}.
		
		\subsubsection{Metrics:}
\begin{figure} [h!]
			\centering
			\includegraphics[width=0.8\textwidth]{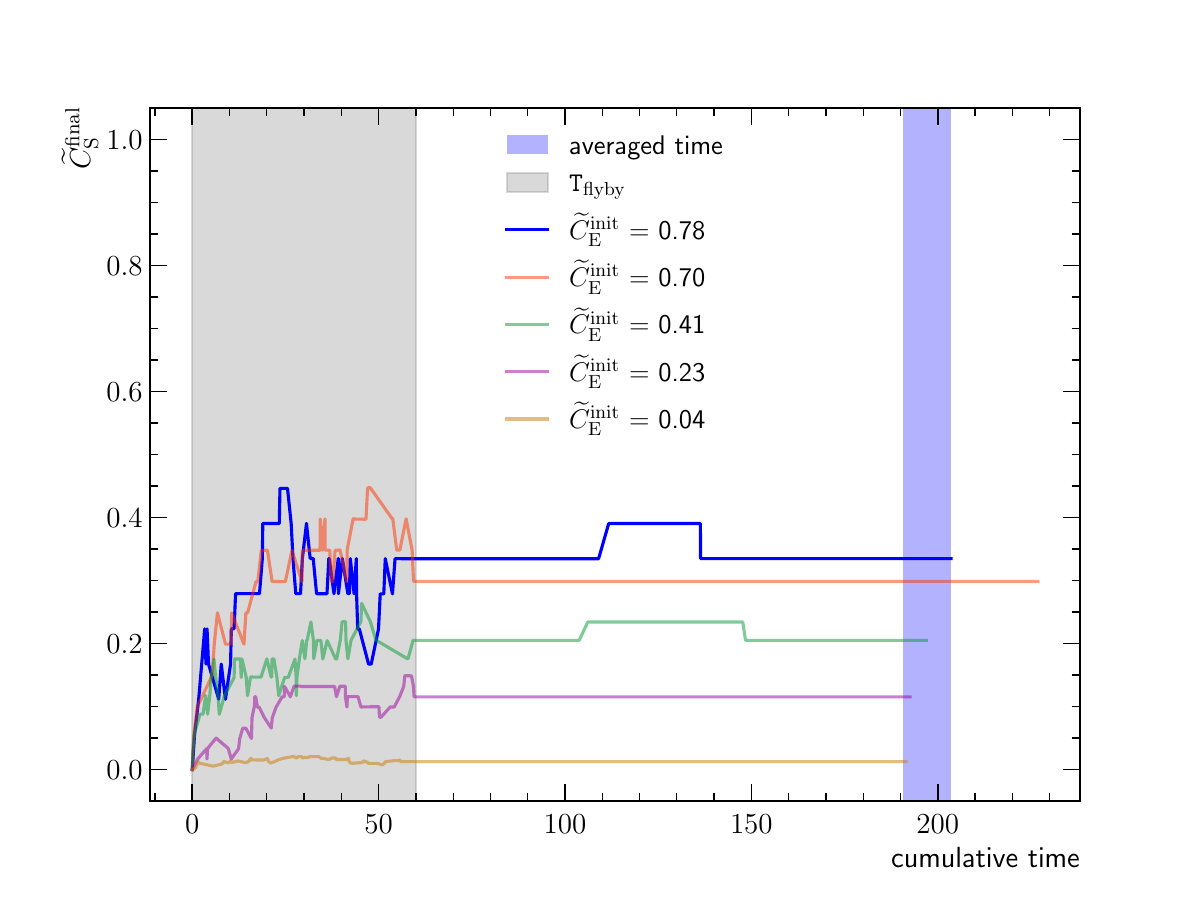}
			\caption{\label{fig:time_behaviour} The time behaviour of five individual runs plotted as a function of time and marked for the different levels of initial \(\widetilde{C}_{E}\). The average time between each interaction is set by \(\mathtt{T}_{int}\) and is an exponentially distributed random variable. The total number of interactions is set by \(\mathtt{N}_{int}\), as a result the average total length is \(\mathtt{N}_{int}\times\mathtt{T}_{int}\). Note, however, that since the time between events is a random variable, the paths above have different lengths. \(\mathtt{T}_{flyby}\) refers to the amount of time for which the signal interacts with the environment marked above in gray; it is another parameter of the dynamics. Notice that even after the flyby time, the environment-environment interactions can give rise to dynamics, as in the blue and green curves. The parameter \(\mathtt{f}_{env}\), set to \(8.2e-3\), is fixed for the above curves. Every reported metric is calculated by averaging over the last 10\% of the interactions, highlighted in blue for the blue curve above.}
		\end{figure}

		We have detailed the metrics used \(C_\alpha\) and \(C_{E}\). We also use the mutual information metric, commonly used in works on quantum Darwinism, such as in the article \cite{riedel-zurek}. We average the metric over the last 10\% of the interactions for each run and report this value. All these features are highlighted in \cref{fig:time_behaviour}.
	\end{appendices}
	
	\section*{Acknowledgements}
	V.J. acknowledges the use of Grammarly (\href{https://www.grammarly.com}{grammarly.com}) as a language improvement tool for this article. V.J. thanks N.B. for helping develop the figures for the present article.
\end{document}